\documentclass[fleqn,usenatbib]{mnras}

\usepackage{newtxtext,newtxmath}

\usepackage[T1]{fontenc}

\DeclareRobustCommand{\VAN}[3]{#2}
\let\VANthebibliography\thebibliography
\def\thebibliography{\DeclareRobustCommand{\VAN}[3]{##3}\VANthebibliography}

\usepackage{graphicx}	
\usepackage{amsmath}	
\usepackage{amssymb}	
\usepackage{siunitx}
\usepackage{pdflscape}
\usepackage[inline]{enumitem}

\title[Inverted Magnetic Fields]{The Evolution of Inverted Magnetic Fields Through the Inner Heliosphere}

\author[A. R. Macneil et al.]{
Allan R.~Macneil,$^{1}$\thanks{E-mail: a.r.macneil@reading.ac.uk}
Mathew J.~Owens,$^{1}$
Robert T.~Wicks,$^{2,3}$
Mike~Lockwood,$^{1}$
\newauthor 
~Sarah N.~Bentley$^{1}$
and Matthew~Lang$^{1}$
\\
$^{1}$Department of Meteorology, University of Reading, Reading, UK\\
$^{2}$Mullard Space Science Laboratory, University College London, Surrey, UK\\
$^{3}$Institute for Risk and Disaster Reduction, University College London, London, UK
}
\date{Accepted 1 April 2020. Received 28 February 2020}

\pubyear{2020}

\begin{document}
\label{firstpage}
\pagerange{\pageref{firstpage}--\pageref{lastpage}}
\maketitle

\begin{abstract}
Local inversions  are often observed in the heliospheric magnetic field (HMF), but their origins and evolution are not yet fully understood.Parker Solar Probe has recently observed rapid, Alfv\'enic, HMF inversions in the inner heliosphere, known as `switchbacks', which have been interpreted as the possible remnants of coronal jets. It has also been suggested that inverted HMF may be produced by near-Sun interchange reconnection; a key process in mechanisms proposed for slow solar wind release. These cases suggest that the source of inverted HMF is near the Sun, and it follows that these inversions would gradually decay and straighten as they propagate out through the heliosphere. Alternatively, HMF inversions could form during solar wind transit, through phenomena such velocity shears, draping over ejecta, or waves and turbulence. Such processes are expected to lead to a qualitatively radial evolution of inverted HMF structures. Using  \textit{Helios} measurements spanning 0.3--\SI{1}{AU}, we examine the occurrence rate of inverted HMF, as well as other magnetic field morphologies, as a function of radial distance $r$, and find that it continually increases. This trend may be explained by inverted HMF observed between 0.3--\SI{1}{AU} being primarily driven by one or more of the above in-transit processes, rather than created at the Sun. We make suggestions as to the relative importance of these different processes based on the evolution of the magnetic field properties associated with inverted HMF. We also explore alternative explanations outside of our suggested driving processes which may lead to the observed trend.
\end{abstract}
\begin{keywords}
Sun: heliosphere, Sun: magnetic fields, Sun: solar wind
\end{keywords}

\section{Introduction}\label{sec:intro}

\subsection{Inverted Heliospheric Magnetic Field}\label{sub:inv}

\begin{figure}
    \centering
    \includegraphics[width = 0.45\textwidth]{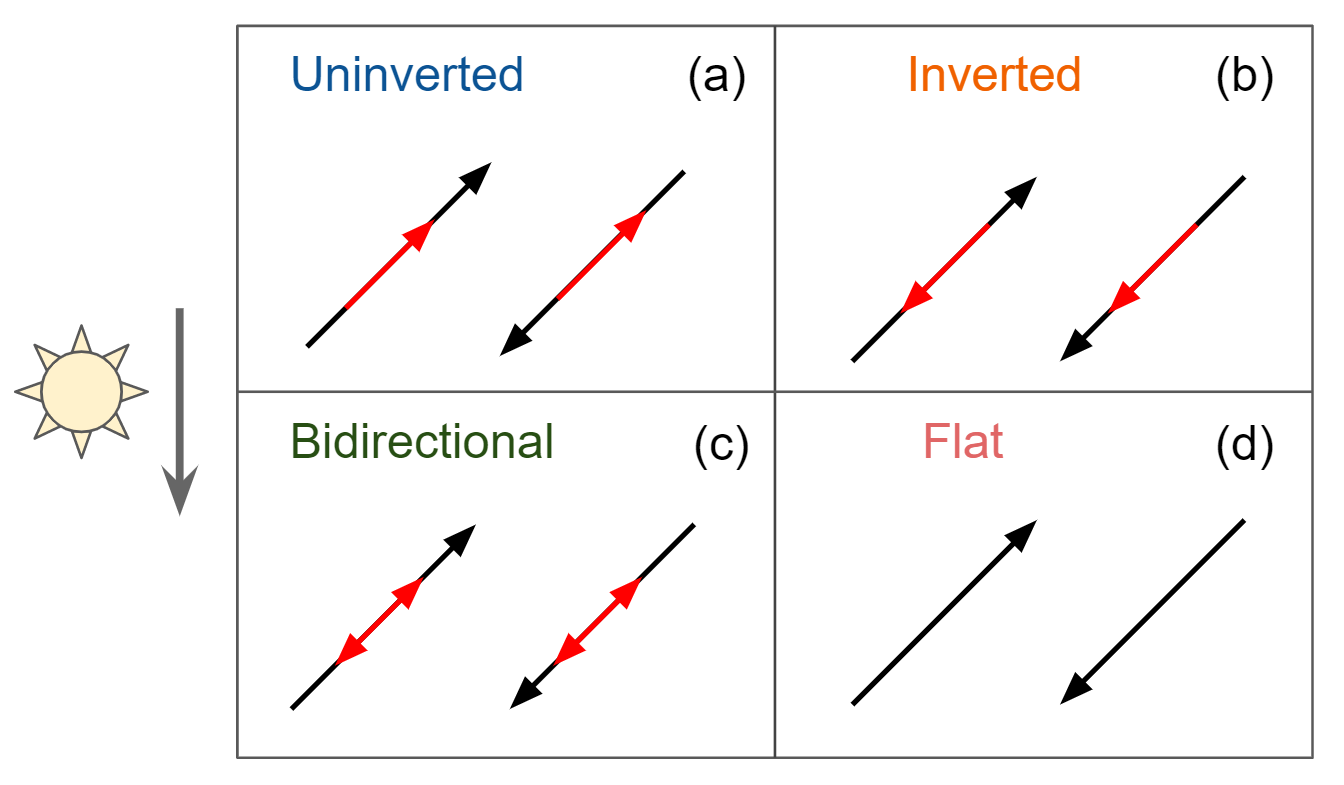}
    \caption{Schematics of the possible local field direction (black arrows) and strahl alignment (red arrows) for 4 possible HMF/strahl configurations: uninverted, inverted, bidirectional strahl, and flat PADs. The HMF is shown in both anti-sunward (on the left of each panel) and sunward (on the right) polarities, and offset from the radial direction due to the Parker spiral.}
    \label{fig:4strahl}
\end{figure}

The heliospheric magnetic field (HMF) regularly exhibits local inversions  (also referred to as reversals)
which are instances where the field is (locally) folded back on itself. These are a subset of  HMF discontinuities \citep[e.g.,][]{Burlaga1969,Mariani1983,Bruno2001,Soding2001}, and are often argued to have their origins in the upstream solar wind or corona through processes typically involving reconnection \citep{Crooker2004,Yamauchi2004,Baker2009b,Owens2013b,Owens2018,Bale2019,Rouillard2020}. As such, they are of much interest in studies on the origins of the solar wind. HMF inversions observed at \SI{1}{AU} have been found to be typically associated with slow solar wind properties \citep{Owens2018}, and have been mapped to sources at separatrices in the corona \citep{Owens2013b}.
Accounting for the presence of inverted HMF has been shown to be a key correction when quantifying total open solar magnetic flux from \textit{in situ} observations \citep{Owens2017}, since inverted field lines can intersect a Sun-centred spherical surface multiple times.
Inversions may be identified through the atypical sunward propagation of the strahl \citep{Kahler1994}; a beam of field-aligned suprathermal electrons which forms in the corona and thus predominantly propagates away from the Sun \citep{Feldman1978,Pierrard2001}. Thus typical strahl orientations are parallel to the field in the anti-sunward HMF sector, and anti-parallel in the sunward sector.
Figures \ref{fig:4strahl}a and b show the strahl directions for uninverted and inverted HMF respectively.

The processes which produce inverted HMF are yet to be identified with certainty. Here we split candidate mechanisms into two groups: 
\begin{enumerate*}
    \item Those which produce inverted HMF structures near the Sun which may then propagate out through the heliosphere, to then decay and straighten out with time.
    \item Those which drive the creation of inverted HMF continually throughout the heliosphere, which thus does not begin to straighten out immediately.
\end{enumerate*}

Short-duration ($\sim$ second), near-incompressible, HMF inversions have been observed in the inner heliosphere by Parker Solar Probe (PSP) \citep{Fox2016,Bale2019,Kasper2019}. These inversions, known as `switchbacks', are characterised by a simultaneous spike in radial velocity on the order of the local Alfv\'en speed, $v_A$, and have been previously found primarily in coronal hole streams \citep[e.g.,][]{Matteini2014}. Switchbacks observed by PSP have been interpreted as outward travelling  Alfv\'enic disturbances, which possibly form at the Sun as a result of coronal jets (placing them into the first of the two groups above). Switchbacks have been previously observed in the inner heliosphere with \textit{Helios} at distances of $\sim \SI{0.3}{AU}$ \citep{Horbury2018}. Inversions of this type are less commonly observed at \SI{1}{AU}, suggesting that they may straighten out, become damped, or otherwise merge into the background solar wind before reaching such distances. 
Modelling of these inversions by \cite{Tenerani2020} has indeed indicated that an inverted magnetic structure, travelling outwards at the Alfv\'en speed, could reach distances accessible to  PSP, but will tend to be removed from the field at greater heliocentric distances as a result of developing density and magnetic fluctuations. 

\begin{figure*}
    \centering
    \includegraphics[width=0.6\textwidth]{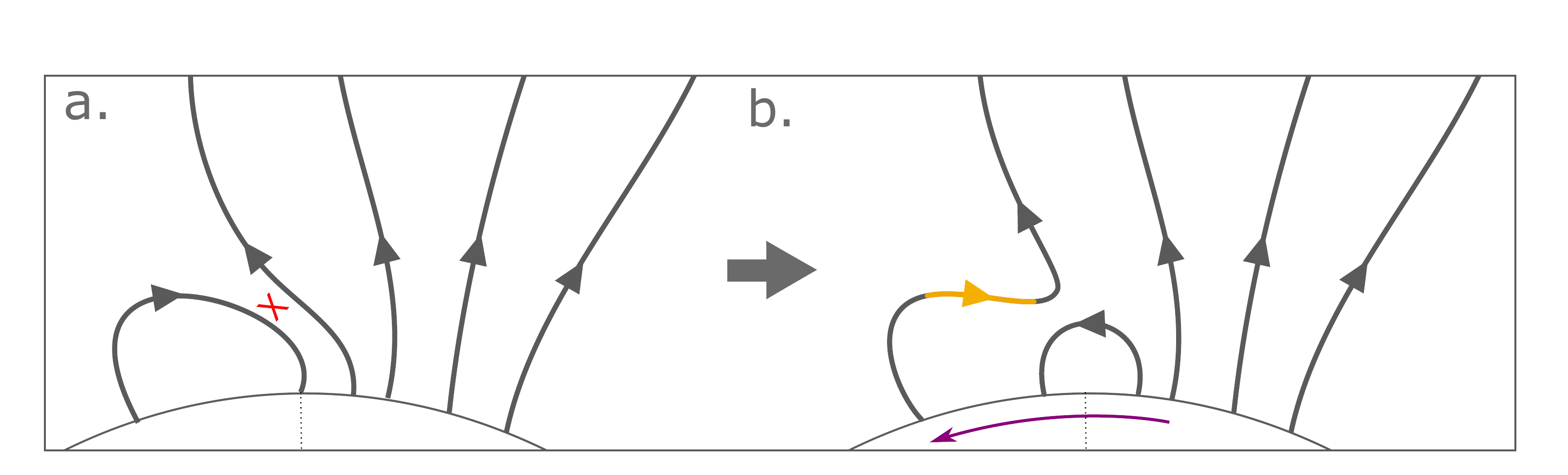}
    \caption{Schematic of interchange reconnection between a closed loop and open magnetic field line in the corona. \textbf{a.} The pre-reconnection configuration, with the reconnection site highlighted by a red `x'. \textbf{b.} The post reconnection configuration, with the kinked portion of the reconnected field line highlighted in orange, and the direction in which the  open field footpoint is transferred is shown with a purple arrow.}
    \label{fig:shearsource}
\end{figure*}

HMF inversions may also be produced close to the Sun following interchange reconnection, as the the opening of a magnetic loop is likely to produce a heavily kinked newly-opened flux tube. This process of interchange reconnection and the subsequent kinked/inverted field is illustrated in Figure \ref{fig:shearsource}  \citep[see also Figure 7 in][]{Owens2013b}. Following its formation, the inverted structure may propagate outwards, but again will likely have some finite lifetime before straightening out or being otherwise damped. Simple scaling arguments (assuming an inversion which convects outward at the background solar wind speed, while straightening out at the local Alfv\'en speed) indicate that non-eruptive loops should not be able to form inversions which survive out to \SI{1}{AU} \citep{Owens2020}. However, as above, the more detailed MHD modelling by \cite{Tenerani2020} suggests that plasma effects may result in inversions surviving further into the heliosphere. Inversions which have been attributed to loop opening have been observed in streamer belt solar wind with PSP near perihelion, where the inversions are most likely to be intact \citep{Rouillard2020}.

\begin{figure*}
    \centering
    \includegraphics[width=\textwidth]{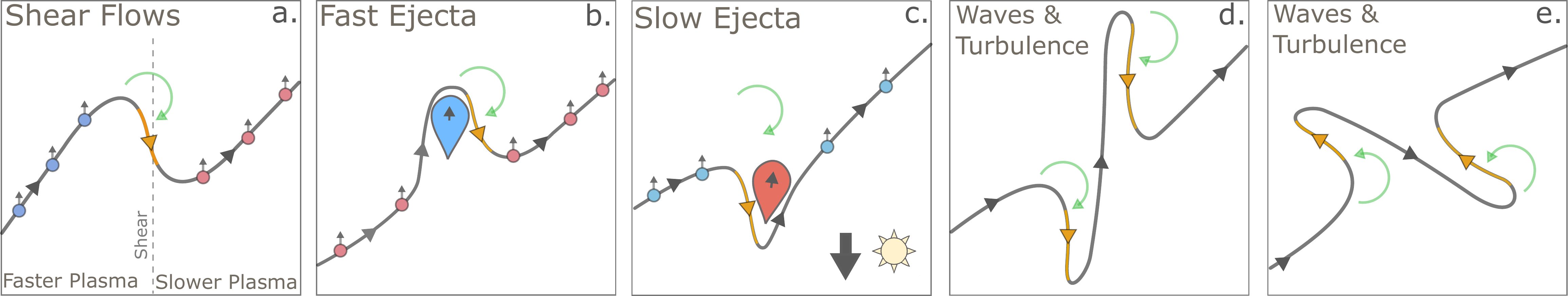}
    \caption{Schematics of processes which may generate inverted HMF near \SI{1}{AU}. Grey lines show HMF which lies in the ecliptic plane and which, if unperturbed, would follow a Parker spiral configuration. Circles with arrows indicate the bulk velocity direction of the solar wind plasma associated with  the field lines. The colour of these circles indicates faster (blue) or slower (red) bulk speed. Orange sections of the field line highlight the inversions relative to the nominal Parker spiral direction.  Green arrows indicate the direction in which the field is deflected relative to the initial Parker spiral direction.
    \textbf{a.}: Inversion through velocity shear. The plane of the shear is indicated by a dashed line. \textbf{b.} (\textbf{c.}): Inversion created by the draping of the field over ejecta  which is propagating faster (slower) than the ambient solar wind. \textbf{d.} and \textbf{e.}: Inversions resulting from fluctuations. }
    \label{fig:shear}
\end{figure*}

We now consider processes which may drive inverted HMF during solar wind transit (the second of the two groups above) which are summarised in Figure \ref{fig:shear}.  \cite{Landi2005,Landi2006} suggested that an inversion could be driven into the field by shear flows in the presence of low-frequency turbulence.
Similarly, \cite{Lockwood2019,Owens2018}  proposed that a velocity shear which is threaded by a magnetic flux tube could lead to a large-scale inversion of the field. They argue that shears might be initially established by the motion of magnetic footpoints at the Sun due to interchange reconnection \citep{Fisk2003}. This motion is illustrated in Figure \ref{fig:shearsource}b, and may constitute a change in the source region for a given flux tube, leading to changes in solar wind properties, such as velocity, along it.

As illustrated in Figure \ref{fig:shear}a, 
velocity shears can drive deflections in the HMF by acting to `stretch' magnetic flux tubes, leading to them increasingly lying in the plane of the shear. In the ecliptic plane, shears in radial flow thus have a maximum possible deflection angle which is reached when the field has been rotated to be radial or anti-radial. 
One consequence of this is that shears in which a slow flow is followed  a faster one  (slow--fast shear) will only drive inverted HMF by rotating the field away from the radial direction, per the example in Figure \ref{fig:shear}a. This is a clockwise rotation when viewed from ecliptic north.
A faster flow followed by a slower one (fast--slow) will deflect the field towards the radial direction (anticlockwise), and so cannot create a deflection of \SI{>90}{\degree} from the Parker spiral direction, and thus cannot invert the field.
Thus, radial velocity shears can only invert an initially Parker spiral field through deflections which are clockwise as viewed from ecliptic north.

An inversion generated by a shear will begin to straighten when the shear no longer exists. However, we note that as the Alfv\'en speed drops with heliocentric distance, $r$, the maximum rate at which this can occur follows suit.
For a shear in radial velocity along a flux tube, we also expect the Parker spiral geometry to result in more effective kinking of the field with greater $r$. This is because the Parker field becomes increasingly orthogonal to the radial shear. Although the maximum rate of kinking is still controlled by the local Alfv\'en speed, at greater $r$ increasingly small shears should become viable to invert the field.

Inverted HMF could also be similarly driven by the draping of HMF around ejecta, ranging from small-scale blobs or jets \citep{Sheeley1997,Kilpua2009} to interplanetary coronal mass ejections \citep[ICMEs,][]{McComas1989}. Inversions generated by this draping can not be classified as switchbacks due to the compression involved at the front of the ejecta.
Small-scale blobs are often observed in the slow solar wind at \SI{1}{AU} \citep[e.g.,][]{Kepko2016}, and  may originate from reconnection at streamer-tops \citep[e.g.,][]{Endeve2004} or other nulls in a complex web of coronal separatrices \citep{Antiochos2011}. 
Due to possible links with coronal reconnection, both ejecta draping and velocity shear-driven inversions appear possibly related to the origins of the solar wind.

Examples of draping are illustrated in Figures \ref{fig:shear}b and c for ejecta which are moving faster and slower than the ambient solar wind respectively.  
Following the same argument as for velocity shear above, if the ejecta are propagating radially into Parker spiral-oriented field, then only clockwise-deflected inversions are possible (as reflected in the figure). Ejecta which expand might produce some inverted HMF through the opposite rotation, although the expansion rate would have to be large relative to its propagation speed.
As for velocity shears, we expect ejecta with smaller differences in velocity to become able to invert the field at greater $r$.
Draping-driven inversions can reasonably persist until the point at which the ejecta is accelerated to the background solar wind speed (and has ceased any internally-driven radial expansion).

Plasma waves and fluctuations can deflect the HMF away from the nominal Parker spiral direction \citep{Burlaga1982}, and may thus also drive locally inverted fields, as shown generally in Figure \ref{fig:shear}d and e. As illustrated in the figure, 
we do not necessarily expect a  bias in deflection direction to apply to inversions driven in this way.
As for shears and draping, it is possible that these fluctuation-driven inversions should become more common, or grow in size, with $r$. 
The continual evolution of solar wind turbulence may  tend to generate more, or stronger, HMF distortions, as a steady state does not appear to be reached until $\gtrsim\SI{4.5}{AU}$ \citep{Roberts2010}. This is consistent with simulations which show the more rapid development of solar wind turbulence in instances where the solar wind flow is not closely aligned with the background magnetic field (i.e., at greater $r$) and solar wind expansion is included \citep{Verdini2015}.
Turbulent fluctuations are also key in the generation of inverted HMF in the simulations of \cite{Landi2006} discussed above. 

\subsection{Closed and Disconnected HMF}
Inversions are examples of HMF morphological features. Other such features exist which can also be probed using \textit{in situ} measurements of magnetic field and strahl, and so are identified incidentally when searching for inverted HMF. 
Counterstreaming/bidirectional strahl is defined as oppositely-directed strahl beams found on the same magnetic flux tube (Figure \ref{fig:4strahl}c). Bidirectional strahl is often attributed to the presence of a closed loop in the heliosphere \citep{Montgomery1974,Gosling1987},  
but is also observed when strahl electrons are back-scattered by features such as interplanetary shocks caused by stream interactions and ICMEs \citep{Gosling1987,Steinberg2005,Skoug2006}. 
We note also that due to strahl scattering \citep[e.g.,][]{Hammond1996,Vocks2005}, not all closed HMF loops will necessarily feature detectable bidirectional strahl \citep[see][]{Owens2007}.

Strahl drop outs (also known as heat flux drop outs) are defined by the absence of a strahl beam, and are thought to occur when neither end of a flux tube in the HMF is connected to the Sun \citep[magnetic disconnection;][]{McComas1989,Lin1992,Pagel2005b}. This results in a suprathermal electron pitch angle distribution (PAD) which has no peak due to the strahl (i.e., it is `flat').
However, flat PADs with no identifiable strahl may also form when strahl electrons simply undergo strong scattering, without magnetic disconnection being necessary \citep[see e.g.,][]{Pagel2005b}. 

\subsection{Radial Occurrence of Inverted HMF Signatures}

In this study we attempt to constrain which processes are responsible for inverted HMF in the heliosphere.
To do so, we measure the occurrence rate of samples of inverted HMF, relative to the other possible morphologies, as a function of $r$, using \textit{Helios} 1 measurements. 
\cite{Macneil2020} (hereafter AM20) noted an apparent increase in inverted HMF occurrence as a function of $r$ using  \textit{Helios} data. However, the aim of AM20 was to study the pitch-angle width of the strahl and thus a significant fraction of the data was omitted as part of the quality control for that purpose. It is possible that the omitted data was not random and thus skewed the inverted HMF occurrence statistics. Here, we treat the same dataset in a more appropriate manner for reliably measuring changes in occurrence, in a statistical study which classifies solar wind samples based on the HMF and strahl.

We expect that if inverted HMF is primarily produced close to the Sun (e.g., through jets or as post-reconnection structures) and then tends to straighten out, then a decreasing occurrence trend will be observed. Conversely, if inverted HMF is driven into the field continually (e.g., by shears, draping, or fluctuations) then an increasing trend will instead be found.
We note that the observed occurrence rates may also be subject to other factors, such as expansion of inverted structures, changes in scale size, or  changes to the cross-sections sampled by \textit{Helios} with $r$. 
Further, this analysis does not distinguish between switchbacks and other types of field reversal.
Nevertheless, knowledge of the overall inverted HMF occurrence is an important constraint on any physical processes which ultimately aim to explain the generation of inverted HMF.

The structure of this paper is as follows.
In Section \ref{sec:datmeth} we describe the  \textit{Helios} 1 data, and our method of classifying HMF types for solar wind samples. Section \ref{sec:results} reports the results of occurrence of each classified HMF/strahl type with $r$. In Section \ref{sec:dis}, we discuss these results, as well as their limitations, and any resulting caveats which apply in our interpretation. We draw conclusions in Section \ref{sec:conc}. We further include three appendices which assess the potential impact of certain aspects of our analysis on the results of the study.

\section{Data and Methodology}\label{sec:datmeth} 
\subsection{ \textit{Helios} Data}
The data used in this study are identical to those used in AM20. Here we provide an overview, and defer to AM20 for a more detailed description. We use data from the  \textit{Helios} 1 spacecraft's Plasma Experiment (E1)  and Magnetic Field Experiment (E2). These data were collected between years 1974--1981, over distances 0.3--1 AU, approximately within the ecliptic plane. Electron measurements were made on a \SI{40}{s} cadence with the E1-I2 electron analyser. We use magnetic field vectors, time-averaged to this cadence, to construct electron pitch angle distributions (PADs) of phase space density. Each \SI{40}{s} measurement constitutes a solar wind sample for which the HMF morphology will later be classified.

The E1-I2 instrument's angular bins all lie in the ecliptic plane, and so to ensure that the strahl is captured in each measurement, we remove all data where $|B_z/\mathbf{B}|>0.156$ (i.e., with a strong out-of-ecliptic component). AM20 further remove data where the HMF azimuthal angle, $\phi_{B}$, lies within one of the gaps between the 8 angular bins of the instrument. We do not take this further step here, as we  do not require that the peak is captured; only that a preferential strahl direction is detectable. In Appendix \ref{sec:nogaps}, we demonstrate the importance of including these data for the purposes of the present study.  

Some data are, however, necessarily excluded from study, including samples where there are no valid HMF or electron measurements, and so no classification can be made. 
A significant portion of the removed data corresponds to an anomalous subset of electron measurements highlighted in AM20. These feature a strong enhancement in flux, and a high degree of isotropy, at suprathermal ($\gtrsim\SI{150}{eV}$) energies.  This degree of enhancement of suprathermal flux creates a clearly separate sub-population where strahl cannot be identified, and is similar to an effect reported due to the use of the  \textit{Helios} high-gain antenna by \cite{Gurnett1977}.
We remove these samples following the procedure of AM20.
For completeness, Appendix \ref{sec:removedplot} displays the occurrence results of this study, with the contributions of both these anomalous flat velocity distribution functions (VDFs) and other periods for which there are missing data included.

\subsection{HMF Classification}\label{sub:class}
We classify each valid sample using combined magnetic field and electron measurements.
First, we take single-energy slices of the electron PADs  at \SI{\sim220}{eV}. If present, the strahl manifests as a  phase space density peak in a given PAD in the parallel or anti-parallel direction. 
The orientation of the strahl for each sample is then classified by identifying peaks in each PAD. A single peak at \SI{0}{\degree} (\SI{180}{\degree}) indicates  parallel (anti-parallel) strahl. Two comparable peaks signifies bidirectional strahl, and the absence of a strong peak indicates a flat PAD. 
For a detailed description of this procedure, see Section 2.1 and Figure 4 of AM20. 

Each sample is further classified using local HMF polarity and strahl alignment. The polarity of the HMF is defined relative to the ideal Parker spiral direction, as calculated based on the heliocentric distance and solar wind bulk speed. Anti-sunward HMF has an azimuthal angle \SI{<90}{\degree} from the Parker angle, while sunward HMF lies \SI{>90}{\degree} from this angle (see the below for more detail). We then classify the HMF as summarised in Figure \ref{fig:4strahl}.
Parallel (anti-parallel) strahl combined with locally anti-sunward (sunward) HMF indicates that the HMF is uninverted. Meanwhile, strahl which is anti-parallel (parallel) to locally anti-sunward (sunward) field indicates locally inverted HMF. The HMF is classified as uninverted or inverted only when the strahl exhibits a monodirectional peak. Samples with bidirectional strahl (possible closed HMF) or flat PADs  (possible disconnected HMF) are represented as their own classes.
Following this classification, we examine radial trends by binning the \SI{40}{s} HMF samples into discrete distance bins. Our choice of number of bins is based on predicted errors in occurrence estimates, and is detailed in Appendix \ref{sub:bins}.

\subsection{HMF Polarity and Orientation}
\begin{figure*}
    \centering
    \includegraphics[width= 0.7\textwidth]{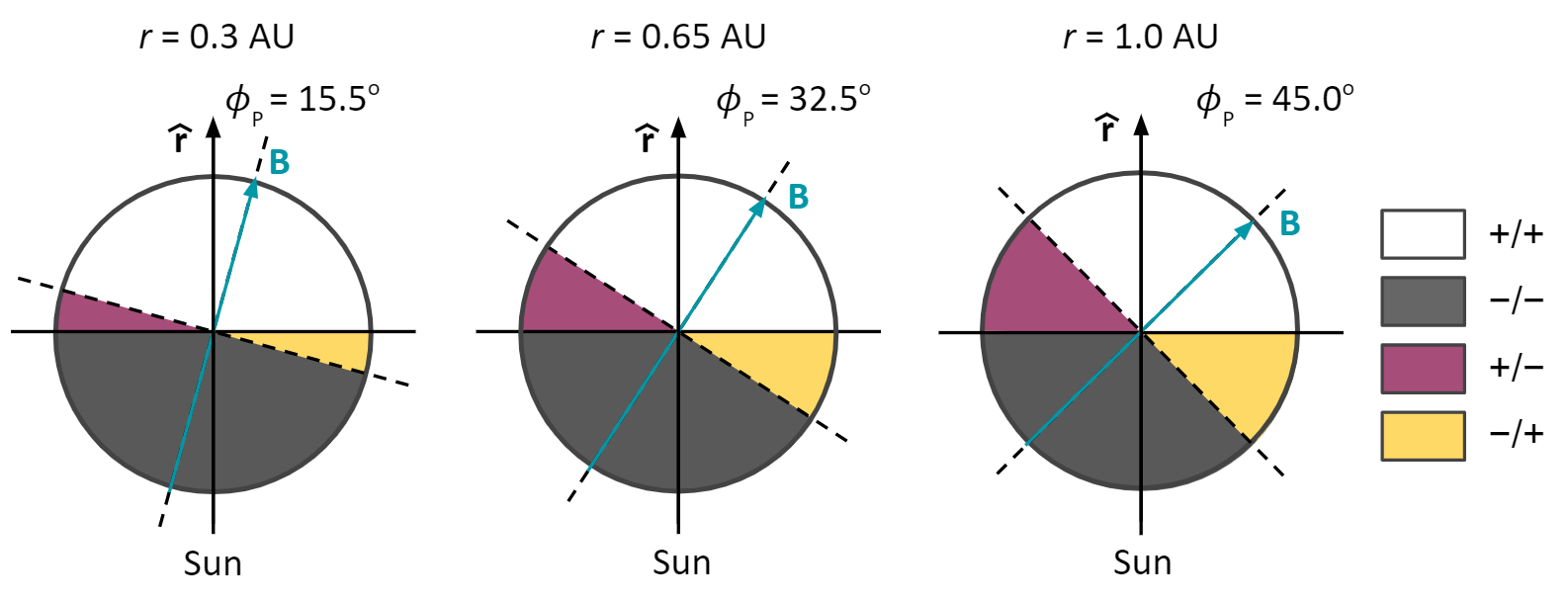}
    \caption{Schematic of the polarity of in-ecliptic Parker spiral HMF at 3 heliocentric distances, calculated for a bulk solar wind speed of \SI{450}{km.s^{-1}}. The magnetic field vector of the Parker spiral magnetic field, $\mathbf{B}$, is shown in cyan. The angular range for which HMF is anti-sunward (sunward) relative to both the radial direction, $\mathbf{\hat{r}}$, and the Parker spiral direction is marked in white (grey) and denoted `$+/+$' (`$-/-$'). The range for which the field is positive (negative) relative to the radial direction and negative (positive) relative to the Parker direction is marked in purple (yellow) and labelled `$+/-$' (`$-/+$').}
    \label{fig:parkerpol}
\end{figure*}
A subtle yet important point of consideration in the above procedure is  the precise definition of HMF polarity. Polarity may be defined relative to the radial direction \citep[i.e., the sign of the radial HMF component, $B_r$, as in][]{Kahler1994,Owens2018}, or relative to the expected mean Parker spiral field $B_P$ \citep[a `Parker' inversion, as used in this study, and by][]{Balogh1999,Heidrich2016,Macneil2020}. 
The difference between these two polarity types is demonstrated in Figure \ref{fig:parkerpol}, which colours azimuthal angular sectors based on their polarity as defined relative to the radial and Parker spiral directions. 
The figure shows that the degree of  overlap between the two polarity types falls with radial distance, as the nominal Parker angle diverges from the radial direction. 

When investigating inverted HMF for the purpose of correcting estimates of total open heliospheric flux \citep[e.g.,][]{Owens2017}, only the radial field component, $B_r$, is considered, and so defining polarity relative to $B_r$ is most suitable. 
Here, we define the polarity instead relative to $B_P$, to ensure that the size of deflection which constitutes an inversion relative to the expected field direction is constant across all distances, and for both directions of deflection (as shown in Figure \ref{fig:parkerpol}). This is crucial as we wish to consistently examine the evolution of inversions with heliocentric distance.

\begin{figure}
    \centering
    \includegraphics[width=0.4\textwidth]{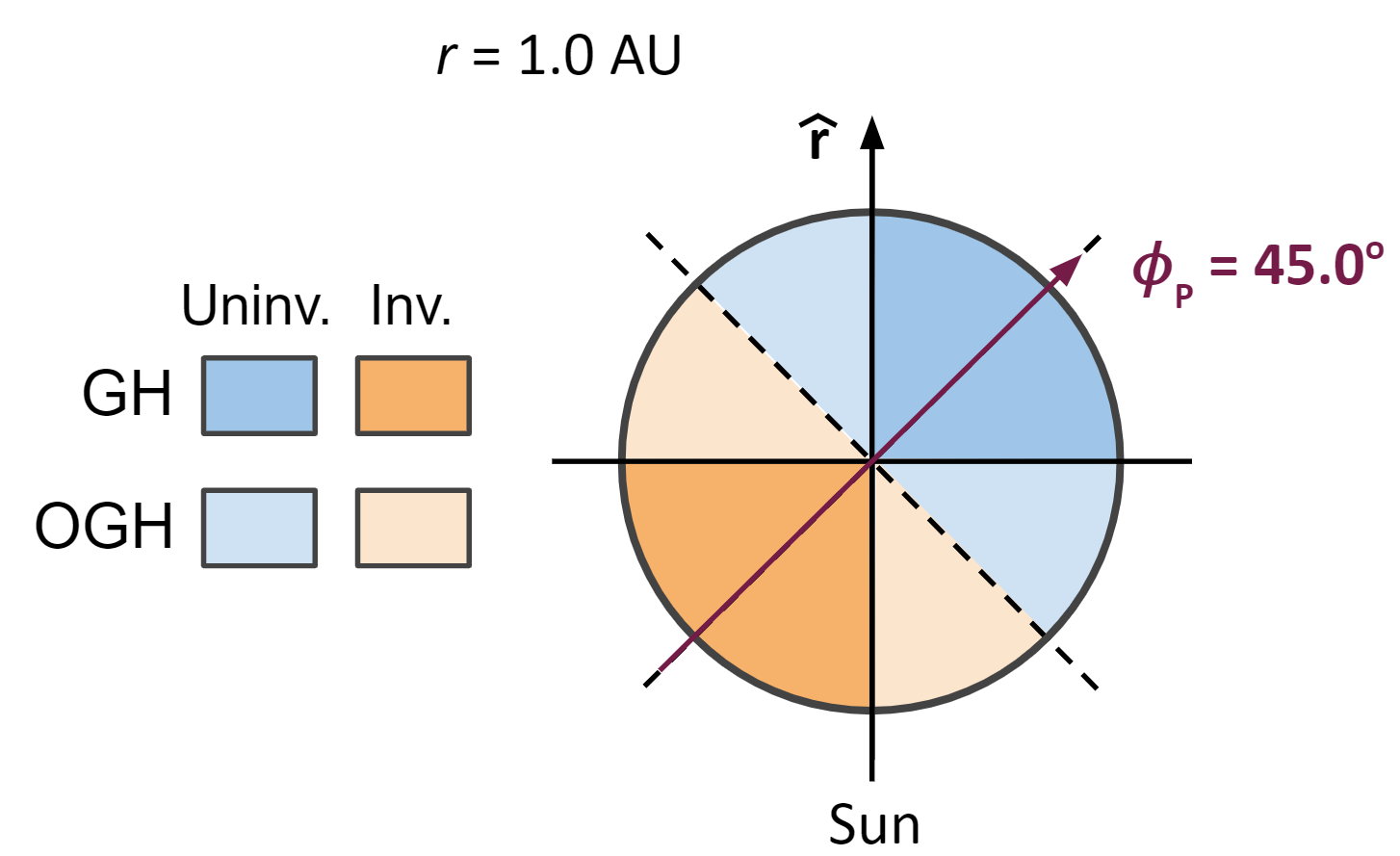}
    \caption{Schematic of HMF azimuthal angles which constitute GH and OGH orientation, relative to a nominal Parker spiral direction of \SI{45}{\degree}. In this example  the nominal field direction is anti-sunward, thus sunward HMF is inverted.}
    \label{fig:OGHschem}
\end{figure}

As an extension of our analysis, we follow \cite{Lockwood2019} in further dividing HMF samples by azimuthal angle $\phi$ into `gardenhose' (GH) and `ortho-gardenhose' (OGH) sectors (so-called because of the orientation of the classical Parker spiral). GH field is that for which $\phi$ falls within \SI{\pm45}{\degree} of the nominal Parker spiral or anti-Parker spiral angle, while OGH field is any other angle. This is shown schematically for HMF at $r\sim\SI{1}{AU}$ in Figure \ref{fig:OGHschem}. HMF which is locally inverted  may fall into either OGH  or GH  sectors, and GH sector inversions represent the largest departure from the unperturbed direction.

\section{Results}\label{sec:results}

\subsection{HMF/Strahl Type Occurrence}\label{sub:occres}
\begin{table*}
\begin{tabular}{lllllllll}
\hline
 & Uninvert & Invert & Double & Flat & \textbf{Valid} & Anom. & Misc. & \textbf{Total} \\ \hline
Samples & 215419 & 9426 & 16775 & 6474 & 248094 & 16343 & 1332 & 265759 \\
\% of Total & 81.0 & 3.5 & 6.3 & 2.4 & 93.4 & 6.1 & 0.5 & 100 \\ \hline
\end{tabular}\caption{Number of samples of `valid' (uninverted, inverted, bidirectional (`double') strahl and flat strahl), and `invalid' (anomalous and miscellaneous) HMF/strahl types as measured by  \textit{Helios} 1. Also shown is the percentage of each type relative to the total number of samples.}\label{tab:numbers}
\end{table*}

Table \ref{tab:numbers} shows the number of samples which belong to each valid HMF/strahl type, and the numbers of samples which are removed due to displaying the anomalous strahl (Section \ref{sec:datmeth}). Also shown are the number of samples discarded as they do not produce a valid classification for other miscellaneous reasons (primarily due to NaN values in the E1-I2 electron data). The total number of samples here refers to those which already meet the criteria described in Section \ref{sub:class}. \SI{93.4}{\percent} of these samples produce a valid HMF/strahl type classification. As expected, uninverted HMF is by far the dominant HMF type.  The  majority of `invalid' classifications are due to the presence of anomalous strahl. In Appendix \ref{sec:removedplot}, we show and discuss the occurrence of these invalid types alongside the valid ones.

\begin{figure*}
    \centering
    \includegraphics[width =1\textwidth]{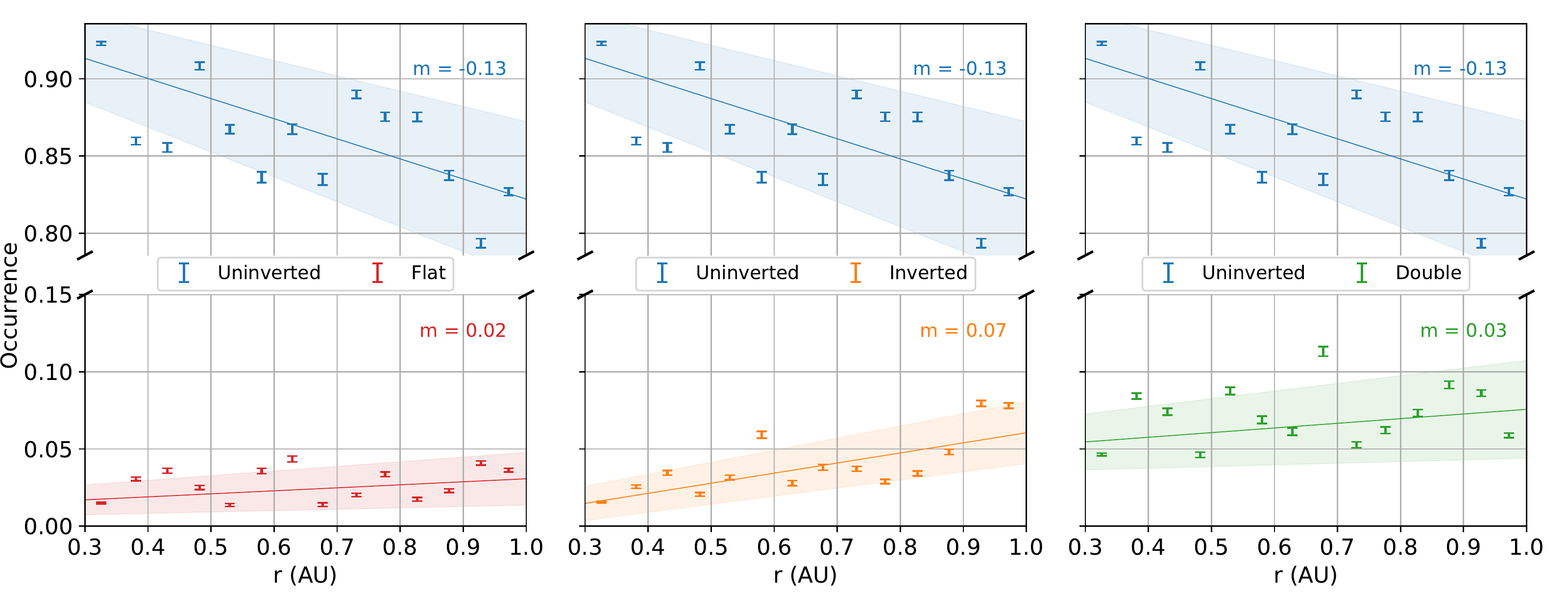}
    \caption{Plots of occurrence for 4 HMF types relative to the ideal Parker spiral direction; uninverted (all plots), no strahl/`flat' (left), inverted (centre), bidirectional strahl/`double' (right) plotted in bins of heliocentric distance $r$.  Error bars are derived from Equation \ref{eq:std}, where the measured occurrence is used as $P\{C\}$ for each HMF/strahl type.  Lines of best fit are shown for each type, with gradient labelled as $m$, and shaded regions corresponding to the bounds calculated using upper and lower fitting errors in both the gradient and intercept. Each plot is split in $y$ to show clearly the trends in both low and high-occurrence types, with the $y$-axis scale identical in both halves.}
    \label{fig:mainres}
\end{figure*}

From Appendix \ref{sub:bins}, we find that an acceptable percentage error of \SI{10}{\percent} in occurrence rate (for classes with occurrence rate $>0.01$) corresponds to $N=\SI{e4}{}$ samples in each bin. This minimum $N$ can be ensured by splitting the  \textit{Helios} 1 data into at most 14 distance bins of equal width ($\sim\SI{0.05}{AU}$). 
Figure \ref{fig:mainres} plots the occurrence of the 4 valid HMF/strahl types in 14 evenly-spaced distance bins, against bin distance $r$.   A line of best fit is calculated for each type, with gradient $m$. 
The occurrence of uninverted HMF falls off with $r$ at the expense of the other valid HMF types.
The linear fit indicates a decrease in the occurrence of uninverted HMF from $\sim0.9$--\SI{0.83}{}.  

Inverted HMF occurrence increases with $r$, from $\sim0.015$--0.065, based on the linear fit between 0.3 and \SI{1}{AU}; a factor of $\sim4$. We note that the two outermost points are outliers from the linear fit, and have occurrence $\sim0.08$. Despite these outliers, the scatter of occurrence about the best fit line is smaller for inverted HMF than for the other plotted types. 

The occurrence rate of HMF with bidirectional strahl (`double' in the figure) is greater than that of inverted HMF. 
Bidirectional strahl increases with $r$, however the gradient is more shallow than for inverted HMF, and the overall increase is small in comparison to the spread in values. Based on the linear fit, the occurrence increases from $\sim0.065$ to 0.08 between 0.3 and \SI{1}{AU}; a factor of only 1.25. Note that an absence of radial trend is within the fit uncertainty.

HMF with flat PADs (no clear strahl) very weakly increases in occurrence with heliocentric distance, with an overall increase of $<0.01$ over 0.3--\SI{1}{AU}. It is also the HMF/strahl type with the lowest occurrence of those considered here. The occurrence of this type increases by a factor of 1.3 between 0.3 and \SI{1}{AU}.

\begin{figure}
    \centering
    \includegraphics[trim={10cm 0 10cm 0},clip,width=0.38\textwidth]{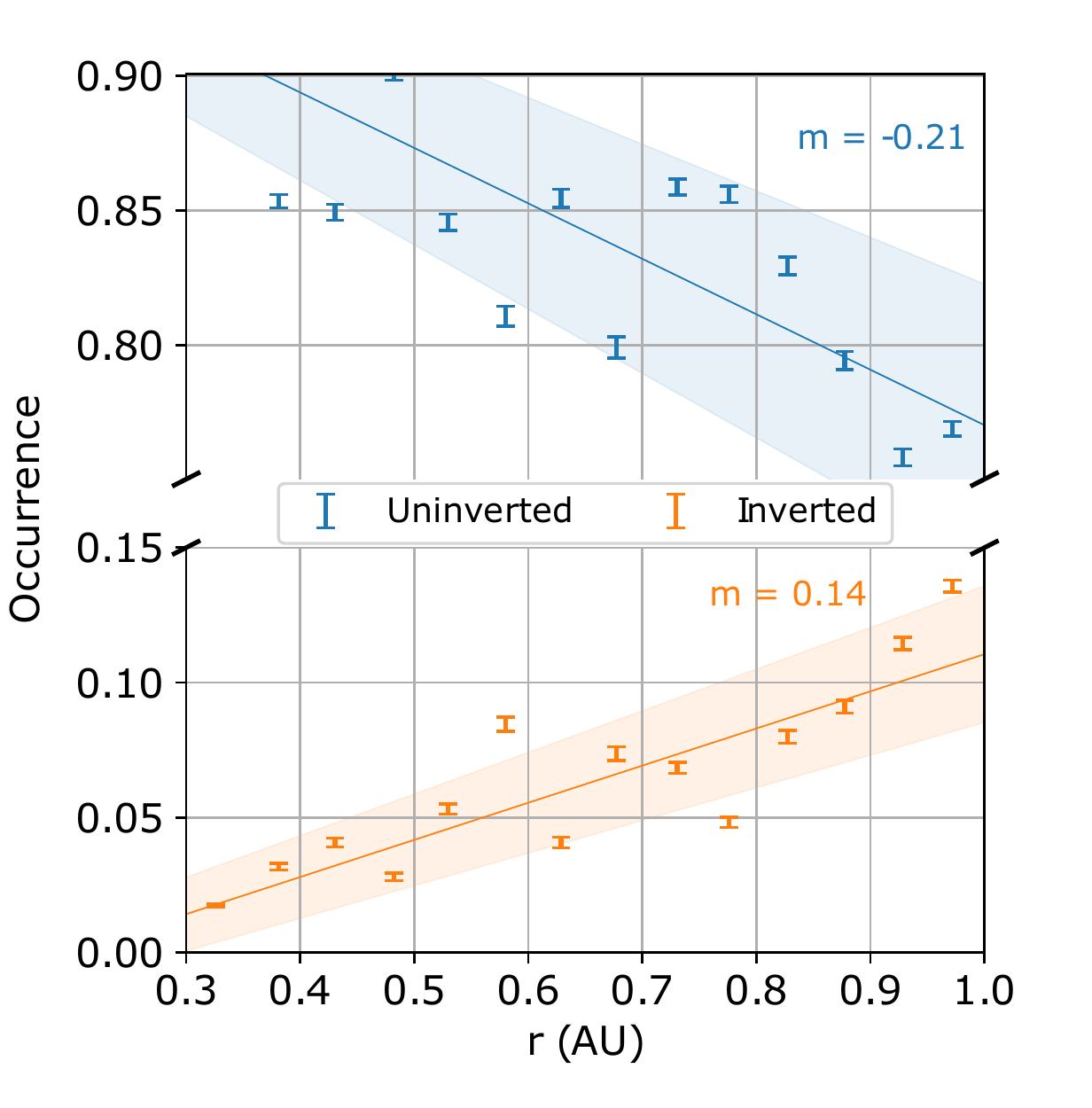}
    \caption{Plot of occurrence for uninverted and inverted HMF in the same format as Figure \ref{fig:mainres}, where results have been computed using HMF polarity defined relative to the radial direction. New best fit lines have been calculated for these results and are also shown.}
    \label{fig:resbr}
\end{figure}
Figure \ref{fig:resbr} plots equivalent results to those of the centre panel of Figure \ref{fig:mainres}, but here the occurrences of uninverted and inverted HMF have been recomputed based on the polarity of the radial HMF component $B_r$, instead of the nominal Parker spiral component $B_P$. (Bidirectional and flat classifications are determined only from the strahl and so are insensitive to HMF polarity and not shown here.) The figure reveals a greater occurrence of inverted HMF at larger $r$ in comparison to that in Figure \ref{fig:mainres} (the positive gradient is around doubled). The occurrence of uninverted flux correspondingly drops-off at an increased rate with $r$. The results are most similar (different) at $\sim\SI{0.3}{AU}$ (\SI{1}{AU}), as expected.

\subsection{HMF Azimuthal Angle}

\begin{figure*}
    \centering
    \includegraphics[width=0.7\textwidth]{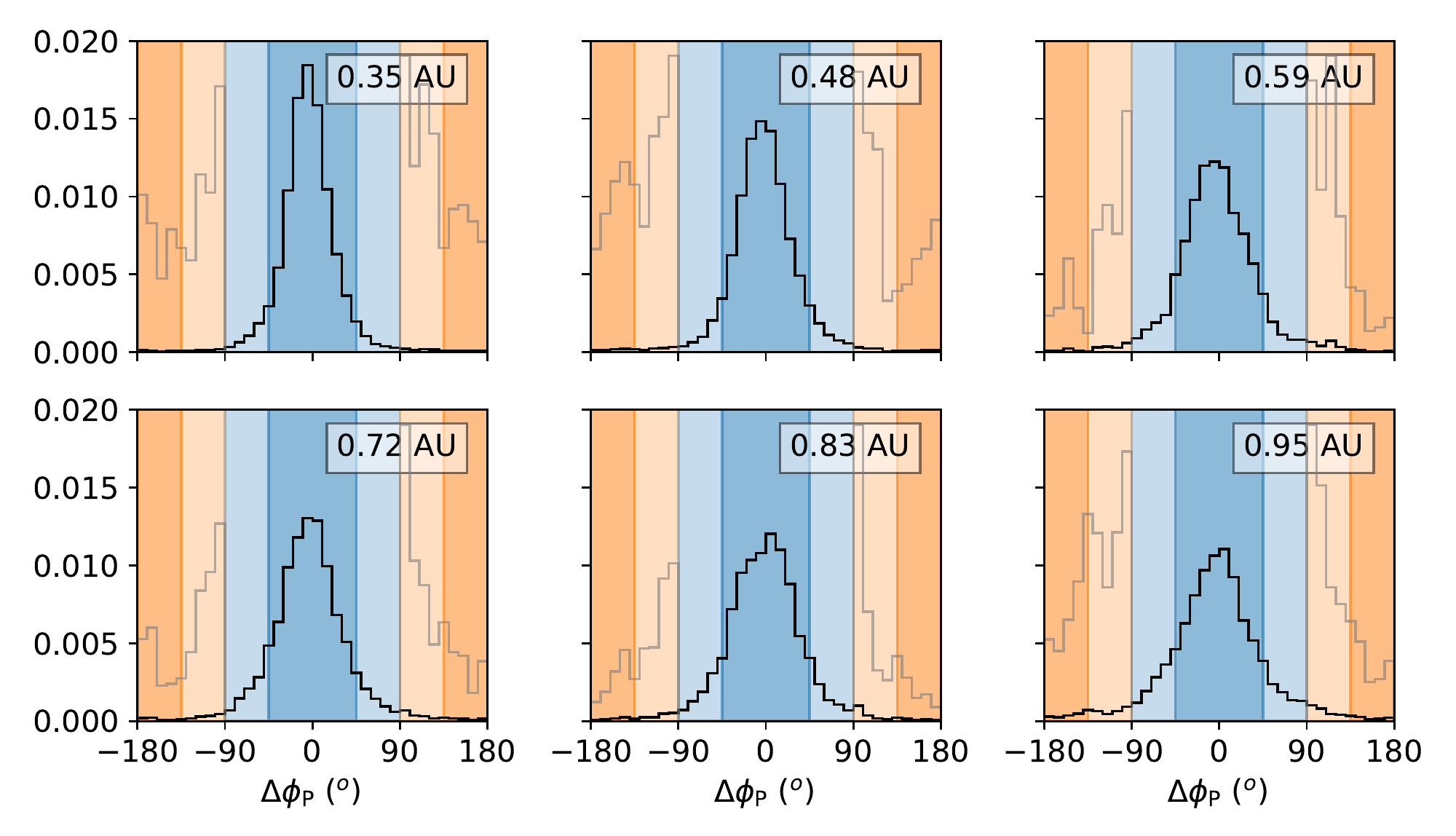}
    \includegraphics[clip,trim = {20.1cm 0 0 0},width=0.17\textwidth]{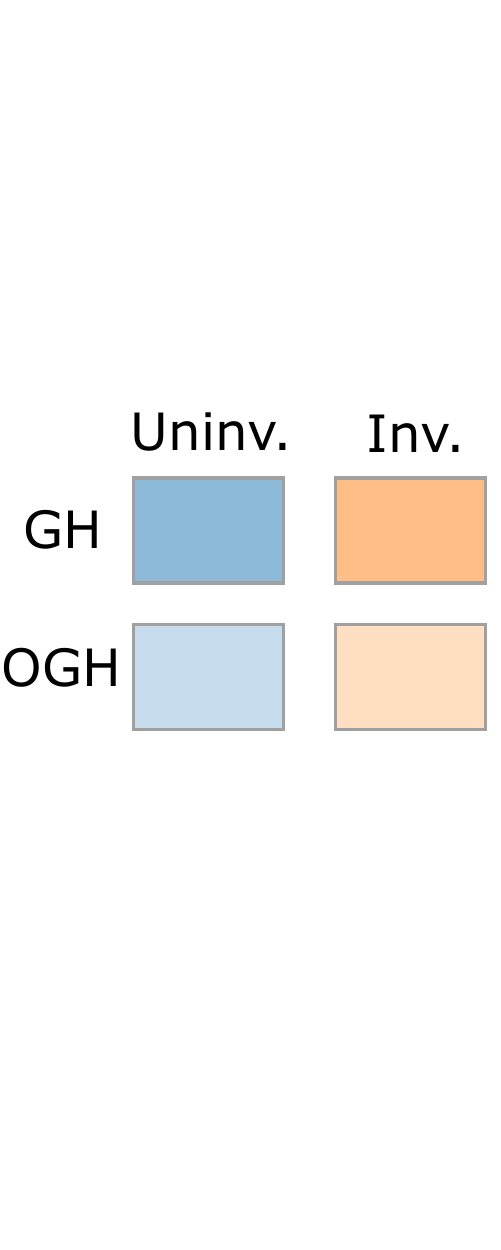}
    \caption{Histograms of the deviation between the observed HMF azimuthal angle from the ideal Parker spiral direction, $\Delta\phi_{\mathrm{P}}$, in 6 radial distance bins. Samples where the strahl is oriented anti-parallel to the local HMF have been shifted \SI{180}{\degree}, such that all points where $|\Delta\phi_{\mathrm{P}}|>\SI{90}{\degree}$ correspond to inverted HMF (see text for details). The sectors are shaded based on GH/OGH and inverted/uninverted angles with the same colour scheme as Figure \ref{fig:OGHschem}. The histograms are normalised such that the area below each line sums to 1. Also plotted in grey is the same histogram, but shown only for inverted HMF, where the plot is zoomed-in to show details of this part of the distribution. The true scales of these sectors are shown by the wings of each histogram.}
    \label{fig:parkerhist}
\end{figure*}

From Section \ref{sec:intro}, solar wind/HMF samples  can be split by the magnetic sector they are expected to belong to, based on the strahl orientation. Parallel (anti-parallel) strahl indicates  anti-sunward (sunward) HMF at the source.  
Figure \ref{fig:parkerhist} plots normalised histograms of the angle $\Delta\phi_{P}$; the difference between the observed azimuthal HMF angle, $\phi$, and the ideal Parker spiral angle, $\phi_P$, in six radial distance bins. Uninverted and inverted HMF are included, but all other classes are excluded.
Six bins are used here in order to ensure sufficient samples to make up each histogram. Based on strahl orientation, all samples from the expected sunward magnetic sector (anti-parallel strahl) have been shifted \SI{180}{\degree}. In this way, the entire distribution is centred around \SI{0}{\degree}, and angles of $|\Delta\phi_{P}| >\SI{90}{\degree}$ are inverted HMF by our prior classification. The plots are coloured to indicate inverted and uninverted HMF, as well as GH and OGH sectors. We also re-plot the inverted HMF components of each histogram in grey, re-normalised so as to highlight the detail.
These distributions, where inverted HMF is separable from uninverted HMF in the opposite magnetic sector, can only be obtained through analysis of the strahl (or another tracer of the source magnetic polarity). They are thus distinct from, and offer additional information to, those presented by \cite{Borovsky2010,Lockwood2019}.

The results of Figure \ref{fig:parkerhist} for the uninverted HMF sectors agree with the results of  \cite{Borovsky2010,Lockwood2019}. 
Nearest the Sun, the angle is most tightly concentrated around the mean value (slightly \SI{<0}{\degree}). At greater $r$, these peaks broaden out, resulting in more samples in the uninverted and inverted OGH sectors. At all distances, the distribution appears relatively continuous across uninverted--inverted, and GH--OGH boundaries.
Near \SI{0.3}{AU}, the small component of inverted HMF which exists is relatively evenly spread between OGH (weakly inverted) and GH (strongly inverted) angles, and weakly favours OGH.
At greater $r$, as the total inverted component increases, the distribution begins to strongly favour OGH angles, indicating that most inverted HMF is only weakly inverted.

\begin{figure}
    \centering
    \includegraphics[width=0.4\textwidth]{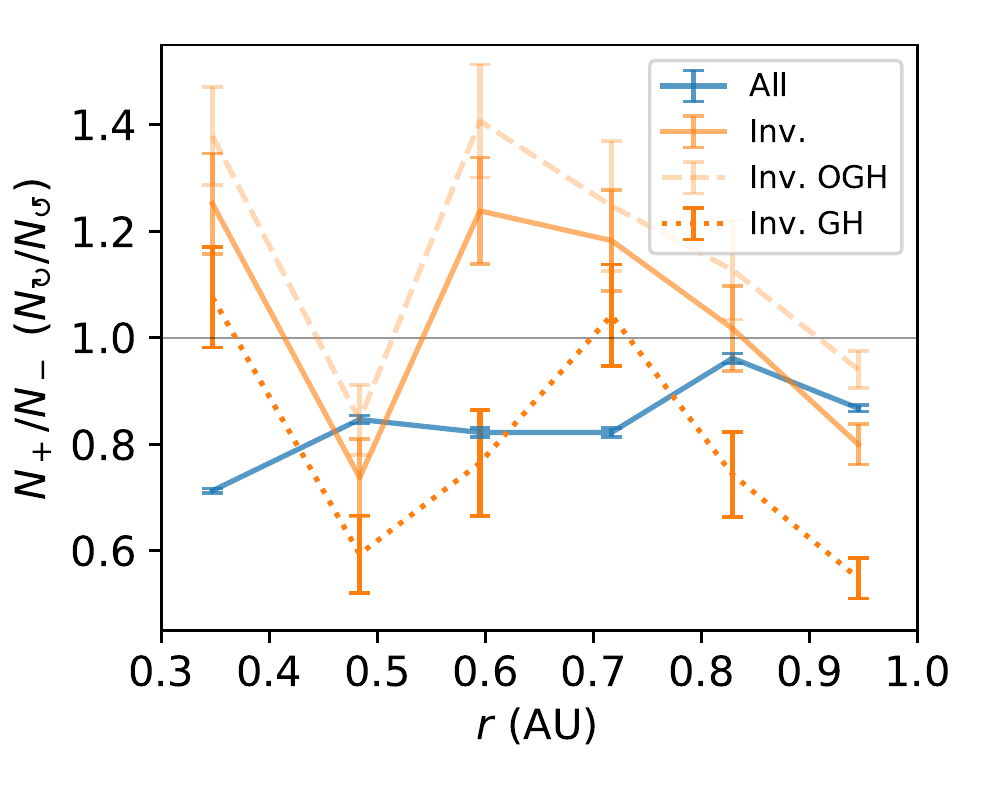}
    \caption{Plot of the ratio of the number of samples where $\Delta\phi_{\mathrm{P}}>\SI{0}{\degree}$ to the number where $\Delta\phi_{\mathrm{P}}<\SI{0}{\degree}$ $(N_+/N_-)$ in Figure \ref{fig:parkerhist} against $r$. The results are shown for all combined inverted and uninverted samples (`All`), for inverted HMF samples only (`Inv.'), and for GH/OGH inverted samples only (`Inv.~OGH`/`Inv.~GH'). Error bars are derived from Equation \ref{eq:std} based on the number of samples in the appropriate class and sector.
    }
    \label{fig:negpos}
\end{figure}

A skew  towards slightly negative values of $\Delta\phi_{\mathrm{P}}$ (an anticlockwise deflection from the Parker angle, Figure \ref{fig:shear}) is present in the distributions of Figure \ref{fig:parkerhist}. This is `underwound' HMF, which skews towards the radial direction, as noted by e.g., \cite{Murphy2002}. We investigate if one direction of deflection is more common, for inverted HMF specifically, in Figure \ref{fig:negpos}. The figure plots $N_+/N_-$, the ratio of the number of positive to the number of negative values of $\Delta\phi_{\mathrm{P}}$  for different sectors of each of the 6 histograms in Figure \ref{fig:parkerhist}, as a function of $r$. 
$N_+/N_- > 1$ $(<1)$ indicates a tendency to clockwise (anticlockwise) deflection, if we assume that there are no deflections  $>\SI{180}{\degree}$. 

$N_+/N_-$ for all samples (with monodirectional strahl) combined is consistently less than one, with a weakly increasing radial trend, which reflects the tendency to underwinding noted above. This tendency is however not present when considering inverted HMF in isolation.
Due to low samples, the ratios for inverted HMF all have large error bars, and are found both above and below $N_+/N_- =1$. $N_+/N_-$ for the OGH sector inversions tends to the greatest value (typically $>1$), while $N_+/N_-$ for the GH sector tends to the lowest (typically $<1$), and $N_+/N_-$ for both combined falls between the two (typically $>1$). 
$N_+/N_-$ for the combined inverted and OGH inverted samples appears to follow a generally decreasing trend, aside from one outlier in the  \SI{0.48}{AU} bin (which exhibits an unusual profile of inverted HMF in Figure \ref{fig:parkerhist}).
However, this trend does not persist clearly when a different number of radial bins is chosen. 
Our interpretation of $N_+/N_-$ for GH-inverted HMF is likely the least reliable of all these ratios, since we expect it to contain the majority of deflections $>\SI{180}{\degree}$, which means that not all samples where $\Delta\phi_{\mathrm{P}}$ is e.g., positive correspond to clockwise deflection.

\begin{figure}
    \centering
    \includegraphics[width = 0.45\textwidth]{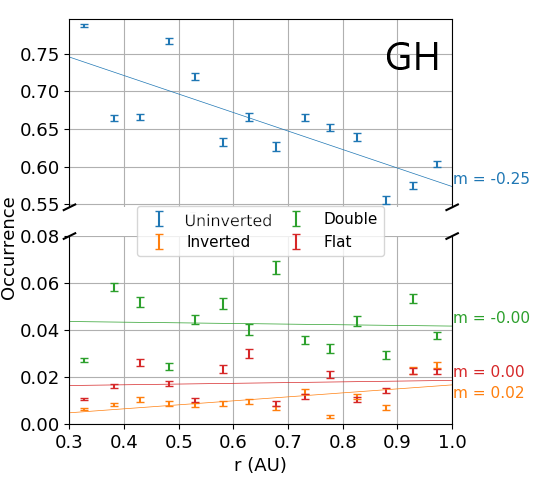}
    \includegraphics[width = 0.45\textwidth]{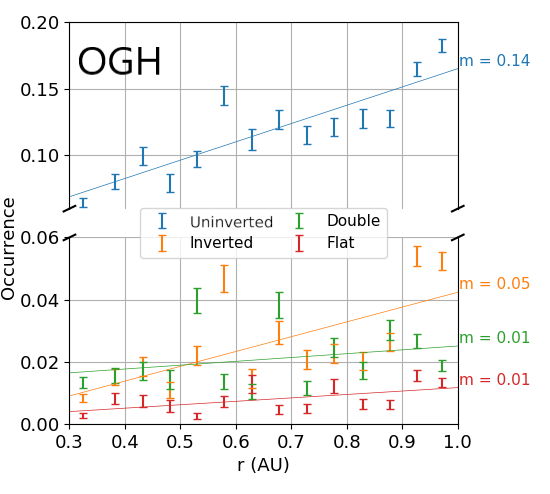}
    \caption{Fractional occurrence of HMF types shown in the same format as Figure \ref{fig:mainres}, but constructed using data only in the GH (top) and OGH (bottom) sectors. The occurrence is calculated as a fraction of all valid HMF/strahl types across all HMF sectors, and so the occurrence does not sum to one in the individual plots. Note the difference in scales in the different portions of each split plot.}
    \label{fig:GHtrends}
\end{figure}

We show the radial trends in occurrence of the 4 HMF/strahl types, split into GH and OGH  sectors (see Section \ref{sec:intro}) in Figure \ref{fig:GHtrends}. 
Uninverted HMF occurrence in the GH sectors drops off more rapidly than that across all sectors combined (in Figure \ref{fig:mainres}), while in the OGH sectors this occurrence increases (this is because of the spreading of HMF angle away from the nominal Parker direction shown in Figure \ref{fig:parkerhist}).
Inverted HMF  increases in both the GH  and OGH sectors, with the primary increase being concentrated in the OGH sectors. The gradients of bidirectional and flat strahl-associated HMF are very weakly positive in both sectors (though show no trend to within uncertainty). The occurrence of both is marginally greater in the GH than OGH sectors (although the total number of samples in GH sectors is far greater than those in OGH sectors). 
\section{Discussion}\label{sec:dis}

\subsection{Limitations and Errors}\label{sub:disccaveats}
There are a number of factors to consider when interpreting results derived from the  \textit{Helios} data set, including data which we discard.
The removal of data with a strong $B_z$ component ($|B_z/\mathbf{B}|>0.156$) may preferentially exclude certain HMF/strahl types. ICME flux ropes, which  often produce out-of-ecliptic field \citep{Burlaga1995}, and tend to exhibit bidirectional strahl, may be particularly affected. 
Table \ref{tab:numbers} and  Appendix \ref{sec:removedplot} show that the occurrence of samples which are excluded because they feature anomalous strahl VDFs (see Section \ref{sec:datmeth}) 
is comparable to that of the HMF types in which we are interested. However, the lack of radial trend in the occurrence of these samples (Figure \ref{fig:fixed}), and the distribution of their associated HMF angles (Figure \ref{fig:parkerhist_rm}), suggest that anomalous strahl is equally likely to occur for all 4 valid types, and so is unlikely to systematically bias our occurrence results. We also find in Appendix \ref{sec:removedplot} that the HMF samples which are discarded due to invalid values in the  \textit{Helios} data have very low occurrence, and so probably do not significantly affect the results.

The removal of samples in this study creates additional gaps in the \textit{Helios} 1 data set, which already has frequent gaps. This excludes the possibility of easily identifying discrete inversion `events', in which the magnetic field can be observed to evolve into an inverted configuration and back. 
Studying these discrete inversions (their frequency, clustering, size,\textit{ etc.}) could yield additional insight which is not available from this analysis. Future missions, particularly PSP and Solar Orbiter \citep{Muller2013}, will provide more continuous data from which inversions can be studied in this way.

The smallest time cadence of the  \textit{Helios} electron data, and thus our HMF classification, is \SI{40}{s}; corresponding to a size of \SI{1.8e4}{km} (or around \SI{3}{R_E}) for a convecting structure of radial velocity \SI{450}{km.s^{-1}}. Structures which are smaller than this are thus not properly represented in our statistics.
The statistical spread of HMF deflections in general are also known to vary in angular size depending on the timescale on which they are examined \citep{Borovsky2008,Lockwood2019}, and so our results are characteristic of the \SI{40}{s} cadence. These factors may be problematic in particular if a significant fraction of the structures in which we are interested are smaller than this minimum detectable scale.

As described in Appendix \ref{sub:bins}, the error bars shown in occurrence estimates are calculated using the expression for a binomial distribution. This assumes an unbiased classification and random sampling of HMF/strahl types by  \textit{Helios} 1.
However, the sampling of these HMF/strahl types is not truly random. The solar wind, and so  HMF structures, are organised into distinct streams and transients, which lead to clustering of HMF/strahl types in time. This clustering will increase the noise about the lines of best fit in radial occurrence, as it allows for more discrepancy  between adjacent bins (e.g., if one more stream associated with a given HMF/strahl type falls within a given radial bin than its neighbours). 
This may explain the particularly strong scatter of bidirectional strahl HMF about the best-fit line in Figure \ref{fig:mainres}, and possible outliers in e.g., Figure \ref{fig:negpos}.
Plotted error bars thus represent only a fraction of the total error, and so it is expected that they do not allow all data to overlap with lines of best fit (if the underlying trends are in fact linear).

\subsection{Inverted HMF Occurrence}\label{disc:main}
We first consider the occurrence of inverted HMF, which is the primary focus of this study. The increase in inverted HMF occurrence in \SI{40}{s} sampled data, (Figure \ref{fig:mainres}) may be representative of the generation and decay of inverted HMF structures. Under this interpretation, while PSP results suggest that some inversions are formed at the Sun \citep{Bale2019,Kasper2019}, there must also be a contribution by some driving process or processes to inverted HMF at 0.3--\SI{1}{AU}. 

Candidate inversion driving processes are summarised in Section \ref{sec:intro}, where we describe how ejecta draping and velocity shears, as well as waves and turbulence,
can produce inversions which  grow or become more common with $r$.
Whatever is the dominant driving process/processes for inversions is likely the same as, or related to, whatever processes are responsible for the general spread observed in HMF azimuthal angle in Figure \ref{fig:parkerhist}. This is because the distributions are largely continuous across the \SI{90}{\degree} cutoff for inverted HMF.

The growth of inverted HMF occurrence is primarily contained in the OGH inverted sectors, and so corresponds to only weakly inverted HMF.
`Strongly' inverted  HMF (GH inverted), is about as common as OGH inversions only in the inner bins of Figure \ref{fig:parkerhist} (i.e., the light grey line is a similar height in the dark and light orange sectors). This inverted HMF is the least likely to be produced as a result of fluctuations, 
and may be part of a subpopulation of inversions which originate close to the Sun. 
If this study were extended down to solar distances observed by PSP, then we might observe inverted HMF occurrence to begin increasing, possibly in the GH sector. This is a possible avenue for future work with PSP.

We argued in Section \ref{sec:intro} that velocity shear and ejecta draping should strongly favour the creation of inversions through deflections in the clockwise direction. In contrast, Figure  \ref{fig:negpos} shows that there is not a strong bias for inverted HMF in either direction. While there appears to be a weak tendency for clockwise deflections, the typical values of $N_+/N_- \sim1.2$ indicate that clockwise deflections make up \SI{\sim54}{\percent} of all inversions. Disregarding ejecta, if velocity shears were responsible for all inverted HMF, then we would expect this value to be \SI{100}{\percent}. 
Based on the weak tendency towards clockwise-deflected inversions,  it appears that velocity shears and ejecta draping in isolation could reasonably drive a minor portion of the increase in inverted HMF samples observed here, and are not dominant over waves and turbulence. 
The existence of a (minor) fraction of inversions for which waves are not responsible is consistent with the observation of inverted structures which are compositionally distinct (and thus of a different solar origin) from their surroundings by \cite{Owens2020}. 

The transverse expansion of solar wind structures can increase the magnitude of HMF deflections which are initially small close to the Sun \citep{Jokipii1989,Borovsky2008,Borovsky2010}, possibly to the point at which the field inverts.
Further, fields which are only deflected to near  $\Delta \phi_P =\SI{\pm90}{\degree}$ by expansion may then be inverted by the effects of one of the other driving processes,  dragging the field into the near-\SI{90}{\degree} inverted sector. 
In simulations by \cite{Squire2020}, expansion has also been found to facilitate the development of initially small Alfv\'enic fluctuations into full field reversals at distances of \SI{35}{R_{\odot}}.
These processes are compatible with the continuous profile of HMF angle across the \SI{90}{\degree} boundary in Figure \ref{fig:parkerhist},  with inverted HMF angles being primarily located in the OGH sector (i.e., close to \SI{90}{\degree}), and with the lack of strong bias towards clockwise or anticlockwise inversions. 

The effects of expansion may somewhat offset the restriction presented in Section  \ref{sec:intro} that shears/draping cannot produce inversions through anticlockwise deflection. If the field already lies in the negative $\Delta\phi_P$ sector (past the radial direction from the Parker spiral), then appropriate shearing could invert the field through anticlockwise deflection, and make a more significant contribution to overall inverted HMF. 
For the above expansion explanation to be valid relies upon small offsets from the Parker spiral angle existing close to the Sun \citep[a reference distance of \SI{5}{R_{\odot}} is used by][]{Borovsky2008}. The processes which we have cited for producing inversions near the Sun in Section 1 (e.g., remnant interchange reconnection structures) are possible sources of these initial offsets from the Parker spiral.

There are other possible explanations for an increasing inverted HMF occurrence with $r$ which do not necessarily require one of the above driving processes. 
Changes to the sizes or dimensions of inverted HMF structures (through e.g., expansion with $r$) might lead to an increasing trend 
without necessarily contributing more inverted flux at each $r$. However, we find that the contribution of inverted HMF to integrated $|B_r|$ and $|B|$ in each radial bin also increases with $r$, suggesting that expansion relative to other structures is not the primary explanation for increased inverted HMF occurrence. Changing dimensions are complicated by the fixed \SI{40}{s} resolution data used here, which might cause a change in characteristic size of inversions  to manifest as a radial trend in occurrence, if that change in size is to or from a scale which we cannot observe.

\begin{figure}
    \centering
    \includegraphics[width=0.45\textwidth]{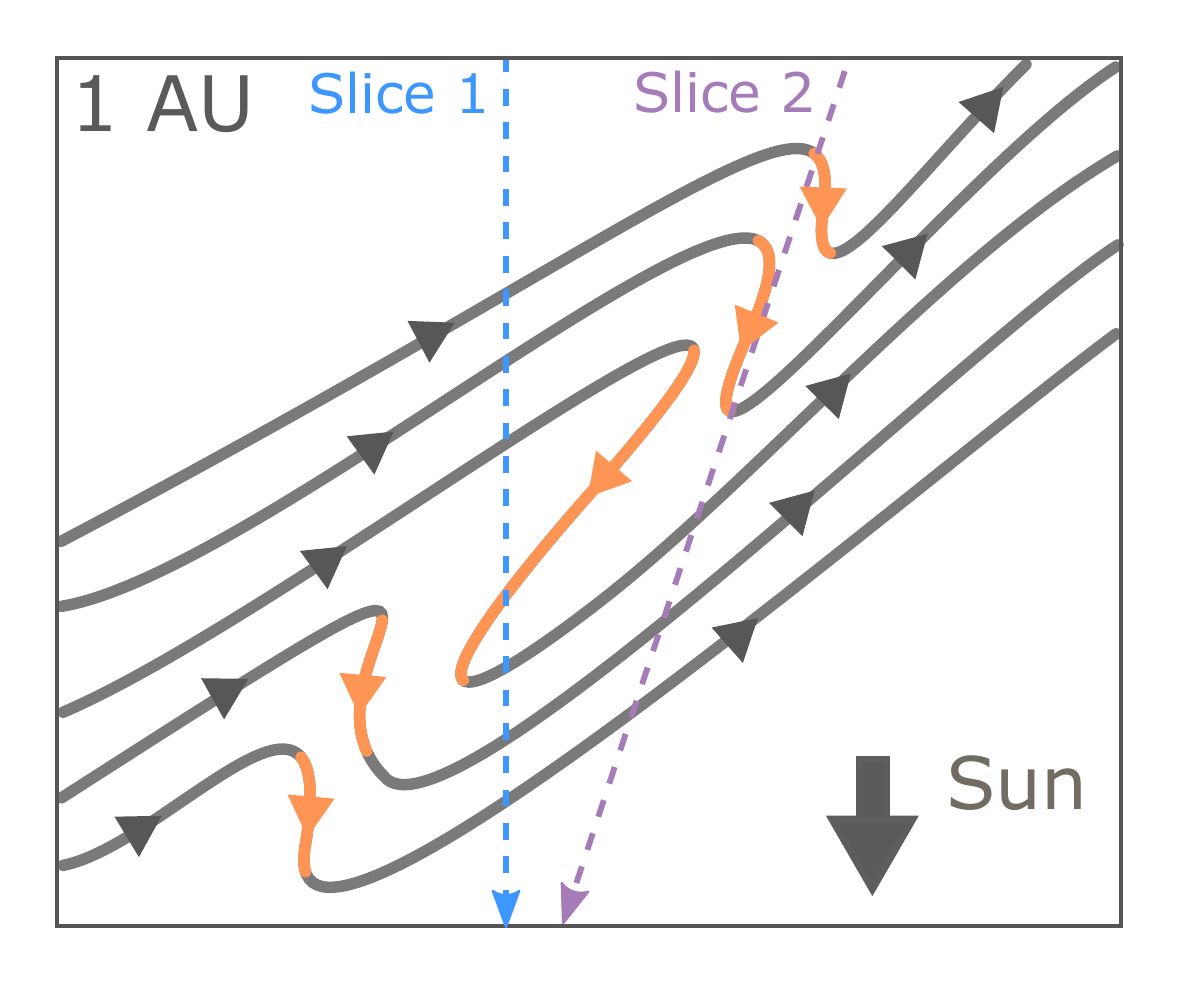}
    \caption{Schematic showing the possible structure of an HMF inversion of finite thickness in two dimensions. The inversion is shown aligned along the Parker spiral direction at $\sim\SI{1}{AU}$. A pair of dashed arrows show the slices taken by \textit{Helios} in the case where the structure is convecting radially (Slice 1) or travelling down the field with some wave speed in the solar wind frame (Slice 2).}
    \label{fig:my_invert}
\end{figure}

The observed occurrence of inverted HMF samples is subject to the path which is followed by \textit{Helios} 1 through inversions as they travel over the spacecraft. 
This is illustrated in two dimensions in Figure \ref{fig:my_invert}. For radially convecting structures, and assuming negligible spacecraft velocity, the spacecraft path is a radial slice  (e.g., Slice 1 in the figure). For  inversions which propagate as a travelling wave down the magnetic field, the slice is at the angle resulting from the sum of the wave velocity and solar wind bulk velocity vectors (e.g., Slice 2 in Figure \ref{fig:my_invert}).
The time spent in the inversion, and so the number of samples, is dependent on the structure's dimensions, orientation, and the crossing speed. Disregarding expansion effects, the typical angle of inversions relative to the radial direction is confirmed to evolve with $r$ by Figure \ref{fig:parkerhist}. Thus, the path taken by \textit{Helios} 1 through the inversions is likely changing with $r$. Whether this should lead to more or fewer inverted HMF samples depends on the dimensions of these structures. The finite thickness of inverted structures may also explain the tendency for more OGH inverted HMF (weakly inverted) to be sampled at greater $r$. The schematic of Figure \ref{fig:my_invert} illustrates that the field is expected to smoothly transition from uninverted to inverted and back, and so a GH (strong) inverted region is likely to be surrounded by OGH inverted flux. 

In summary, the  increasing trend of inverted HMF occurrence with $r$ is likely the result of dynamic processes. These may be active driving of the inversions, or possibly the stretching and rotating of the field which occurs as it expands.
Future detailed explanations of inverted HMF generation must be able to account for these observations of occurrence, as well as the information regarding possible roles of different processes.
The results of this paper thus provide a useful constraint for the interpretation of future observational work, and for the development of models.

\subsection{Flat and Bidirectional Strahl Occurrence}
From Figure \ref{fig:mainres}, there is not a strong trend in occurrence of either bidirectional strahl or flat PAD samples with $r$. If HMF dropouts are primarily responsible for flat PAD samples, then an overall increasing trend with $r$ is expected, since an observer at distance $r$ remains connected to disconnected flux for longer as $r$ increases \citep[see modelling by][]{Owens2007}. 
Similarly, flat PADs resulting from high strahl scattering should also result in an increasing trend, since strahl electrons appear to be continually scattered in the heliosphere \citep[e.g.,][]{Hammond1996}.
The weakness of the observed trend might be because the rate of disconnection, or of strahl becoming fully scattered, is low overall. Alternatively, there may be some contribution from the effects relating to radial evolution, and the interpretation of occurrence results in general, described for inverted flux structures above.

Bidirectional strahl can arise from both closed HMF and strahl back-scattering. The high overall occurrence in bidirectional strahl, compared to inverted HMF, is a result which is highly sensitive to the choice of thresholds when classifying electron PADs as possessing two peaks. However, the lack of an overall radial trend in this occurrence is not. If closed HMF is the primary cause of bidirectional strahl, then we would in fact expect a decreasing radial trend, which is not observed. A large fraction of bidirectional strahl samples being due to back-scatter might explain this, as structures such as shocks (or perhaps HMF inversions themselves) which are associated with this scattering will develop as the solar wind expands. 
\begin{figure}
    \centering
    \includegraphics[clip,trim={4cm 4.8cm 0cm 2cm},width=0.4\textwidth]{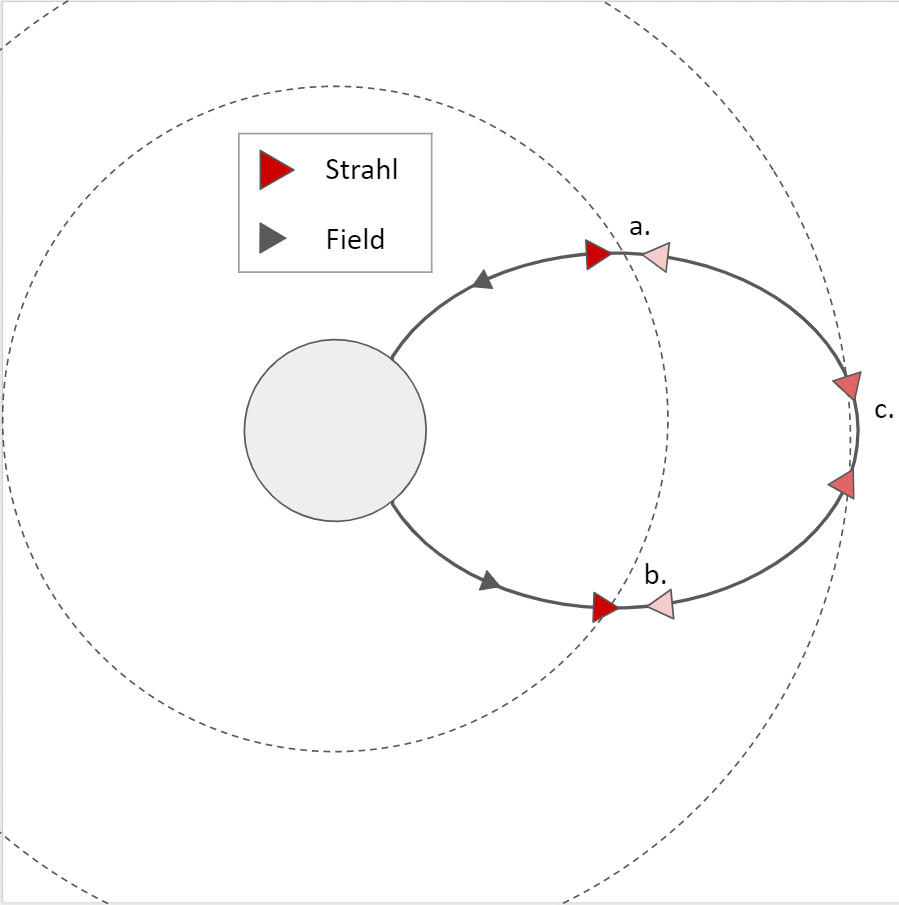}
    \caption{Schematic of an HMF loop and strahl direction along it at 3 locations (a, b, and c) in the heliosphere. The degree of red shading indicates the strength of the strahl at each location, if scattering of strahl is proportional to distance travelled along the loop. Dashed lines represent circles centred on the Sun.}
    \label{fig:loopschem}
\end{figure}

Possible biases in the data might also explain the non-decreasing, highly scattered, bidirectional strahl trend. First, samples of ICME flux ropes, which are often associated with bidirectional strahl, are likely to be under-represented in this study, due to the exclusion of out-of-ecliptic HMF periods. Further, there may be a tendency to observe bidirectional strahl on closed HMF preferentially at greater $r$ due to the scattering of electrons with distance along the field. Figure \ref{fig:loopschem} illustrates a closed loop, and the strength of the counterstreaming strahl at different lengths along it. The clearest bidirectional signatures (two equally intense beams) are expected at the apex of the loop (point c.) where similar scattering is applied to each beam. At either of the legs (points a.~and b.), one beam is expected to be  far less scattered than the other, and so the strahl may appear mono-directional, and the sample will be classified as uninverted HMF. For a given ICME with nose outside of \SI{1}{AU}, \textit{Helios} will observe closer to the base of one leg than a spacecraft near \SI{1}{AU}. Thus, closer to the Sun, a greater fraction of closed HMF may be misidentified as uninverted HMF, leading to the observed trend. Finally, we note that the arguments made regarding expansion and orientation effects for inverted HMF above  may also have some influence here.

\subsection{Radial Inversions}\label{disc:pol}

Any differences between the trend in inverted HMF occurrence when it is calculated by defining HMF polarity relative to the radial direction (as opposed to the Parker direction) is due to the Parker angle straying further from radial at greater $r$, as shown in Figure \ref{fig:parkerpol}.
Inverted HMF occurrence increases when polarity is defined in this way because the $-/+$ sector in Figure \ref{fig:parkerpol} is located nearer to the mean HMF angle than the $+/-$ sector is. Figure \ref{fig:parkerhist} shows that the distribution of HMF azimuthal angles spreads almost symmetrically about the Parker spiral with $r$ (apart from slight deviations show in Figure \ref{fig:negpos}), and so more samples fall into the sector closest to the mean.

The quantification of open solar flux (i.e., flux that threads the solar wind source surface) on the basis of measurements far from the Sun employs radially-inverted HMF measurements as a corrective quantity which is subtracted from the final estimate. Figure \ref{fig:parkerhist} shows that this correction strongly increases with $r$.  \cite{Owens2008} found an increase in  estimated total unsigned heliospheric flux with $r$, which can be partially explained by this increase in inverted HMF which is not being accounted for.
In terms of accurately correcting for inverted HMF, it appears best to produce estimates of total open solar flux using observations made as close to the Sun (or at least as close to \SI{0.3}{AU}) as possible, where the occurrence of inverted HMF is low, and thus the impact of any uncertainty in its estimation is minimised.
However, estimates made by integrating data over numerous highly-eccentric, rapid, orbits -- such as Helios, PSP, Solar Orbiter   --  will have further difficulty in correcting for inverted HMF, due to its strong radial dependence revealed in this study.

\section{Conclusions}\label{sec:conc}

In this study, we have found that the occurrence of inverted HMF (relative to the ideal Parker spiral direction) tends to increase with heliocentric distance $r$ between 0.3 and \SI{1}{AU}.  Inverted HMF is primarily found at azimuthal angles close to \SI{90}{\degree} from the Parker spiral direction, with a minor component more strongly inverted to angles nearer \SI{180}{\degree} which is most significant nearest the Sun. Inversions have a possible bias towards anti-radial (clockwise) deflections from the Parker spiral in most distance bins. These results represent constraints for future studies on inverted HMF generation.

We offer the interpretation that inverted HMF, observed between 0.3--\SI{1}{AU}, is being primarily generated by some continual driving process or processes in the solar wind, rather than being purely a remnant of some processes near the Sun, such as jets or post-reconnection kinks. Inversions generated at the Sun are expected to decay with $r$, and so may still primarily represent inverted HMF observed at distances $< \SI{0.3}{AU}$ by PSP.  

Possible inversion-driving processes include bending of the field by velocity shears along flux tubes, draping over ejecta, or the distortion of the field by waves and turbulence. 
The existence of a significant portion of samples which are inverted in the anticlockwise direction initially suggests that waves and turbulence might be the dominant process in overall contribution to inverted HMF samples, particularly those found near \SI{90}{\degree} from the Parker spiral at greater $r$. 
However, when we consider the effects of expansion on flux tube orientation, subject to an initial offset near the Sun,  a greater contribution from shears and draping becomes permissible.
Shears and draping might therefore be of comparable importance to waves and turbulence, depending on the initial distribution of HMF angles close to the Sun. 
A more reliable identification of which driving processes are dominant will require analysis of the plasma properties associated with inversions, and how these evolve with $r$. We intend to investigate this in a future study.

The driving interpretation of these results in general is subject to the caveats outlined in Section \ref{sub:disccaveats}. Furthermore, alternative explanations for increased inverted HMF occurrence, such as the effects of inversion scale size, expansion, orientation, and three-dimensional structure cannot here be ruled out. However, these possibilities are all still \textit{in situ} dynamical effects.
The different possible interpretations  highlight that a full understanding of HMF morphology with radial distance is not straightforward to obtain.
However, new high quality inner heliosphere \textit{in situ} data are beginning to be returned by Parker Solar Probe \citep{Fox2016}, and soon by Solar Orbiter, for which many of the limitations discussed in Section \ref{sub:disccaveats} will not apply. The extension of this study using both of these missions  (allowing full-sky electron measurements, improved measurement cadence, and greater radial and latitudinal coverage) will allow for this initial insight gained from the  \textit{Helios} mission to be capitalised upon.

\section*{Acknowledgements}
Work was part-funded by Science and Technology Facilities Council (STFC) grant No.\ ST/R000921/1, and Natural Environment Research Council (NERC) grant No.\ NE/P016928/1. 
RTW is supported by STFC Grant ST/S000240/1.
We acknowledge all members of the  \textit{Helios} data archive team\footnote{http://helios-data.ssl.berkeley.edu/team-members/} who made the  \textit{Helios} data publicly available to the space physics community. We thank David Stansby for making available the new  \textit{Helios} proton core data set\footnote{https://doi.org/10.5281/zenodo.891405}.
This research made use of Astropy,\footnote{http://www.astropy.org} a community-developed core Python package for Astronomy \citep{astropy:2013,astropy2018}. This research made use of HelioPy, a community-developed Python package for space physics \citep{Stansby2019}.
Figures besides \ref{fig:4strahl}--\ref{fig:OGHschem},  \ref{fig:my_invert}, and  \ref{fig:loopschem} were produced using the Matplotlib plotting library for Python \citep{Hunter2007}. This work was discussed at the ESA Solar Wind Electron Workshop which was supported by the Faculty of the European Space Astronomy Centre (ESAC).

\bibliographystyle{mnras}
\bibliography{switchback} 



\appendix

\section{Treatment of Gaps in Azimuthal Coverage}\label{sec:nogaps}
\begin{figure*}
    \centering
    \includegraphics[width=.7\textwidth]{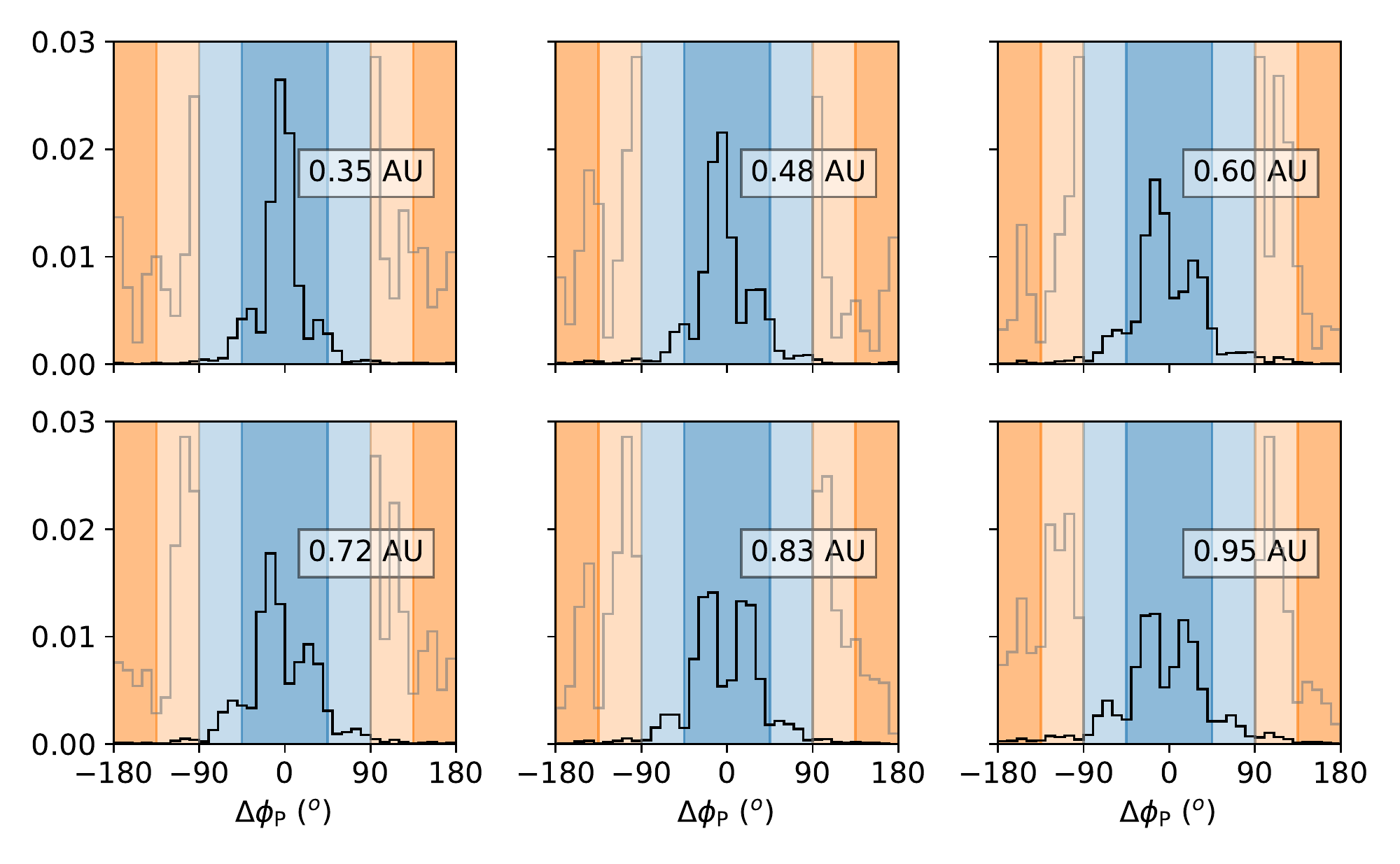}
    \includegraphics[clip,trim = {20.1cm 0 0 0},width=0.185\textwidth]{angle_spread_Parker_colours.pdf}
    \caption{Normalised histograms of HMF angle in the same format as Figure \ref{fig:parkerhist}. The data used to generate these histograms excludes samples where the HMF azimuthal angle does not fall within one of the  \textit{Helios} 1 E1-I2 angular bins.}
    \label{fig:parkerhist_gaps}
\end{figure*}
Section \ref{sec:datmeth} describes how data where the HMF azimuthal angle does not align with one of the 8 detector angular bins were included in this study, in contrast to the analysis of AM20. Figure \ref{fig:parkerhist_gaps} demonstrates the impact of this choice, by re-plotting Figure \ref{fig:parkerpol}, but with data points where the HMF lies outside of the E1-I2 angular bins discarded. In bins near \SI{1}{AU}, where the nominal Parker spiral direction is $\sim \SI{45}{degree}$, there are large notches near the centres of the histogram peaks; particularly for uninverted HMF.

The angular bins in E1-I2 coverage, and so too the gaps, are spaced \SI{45}{\degree} apart in azimuth.  The gaps between the angular bins are centred at $\sim \SI{0}{\degree}$,  $\pm\SI{45}{\degree}$,  $\pm\SI{90}{\degree}$,  $\pm\SI{135}{\degree}$, and \SI{180}{\degree}, with some slight variability of $\sim\SI{2.5}{\degree}$. The notches in peak uninverted HMF occurrence near \SI{1}{AU} are thus the result of  the gaps aligning with mean the Parker spiral angle. 

Comparing to Figure \ref{fig:parkerhist_gaps}, we see that the peaks of occurrence of uninverted HMF around the Parker angles in Figure \ref{fig:parkerhist} appear intact. Thus it is unlikely that the peak of the strahl beam falling between the E1-I2 angular bins is resulting in a large fraction of missed events, when we do not explicitly exclude them. Removing the data which corresponds to gaps in the electron analyser thus excludes a large volume of data unnecessarily.

\section{Removed Data}\label{sec:removedplot}

\begin{figure}
    \centering
    \includegraphics[width=0.45\textwidth]{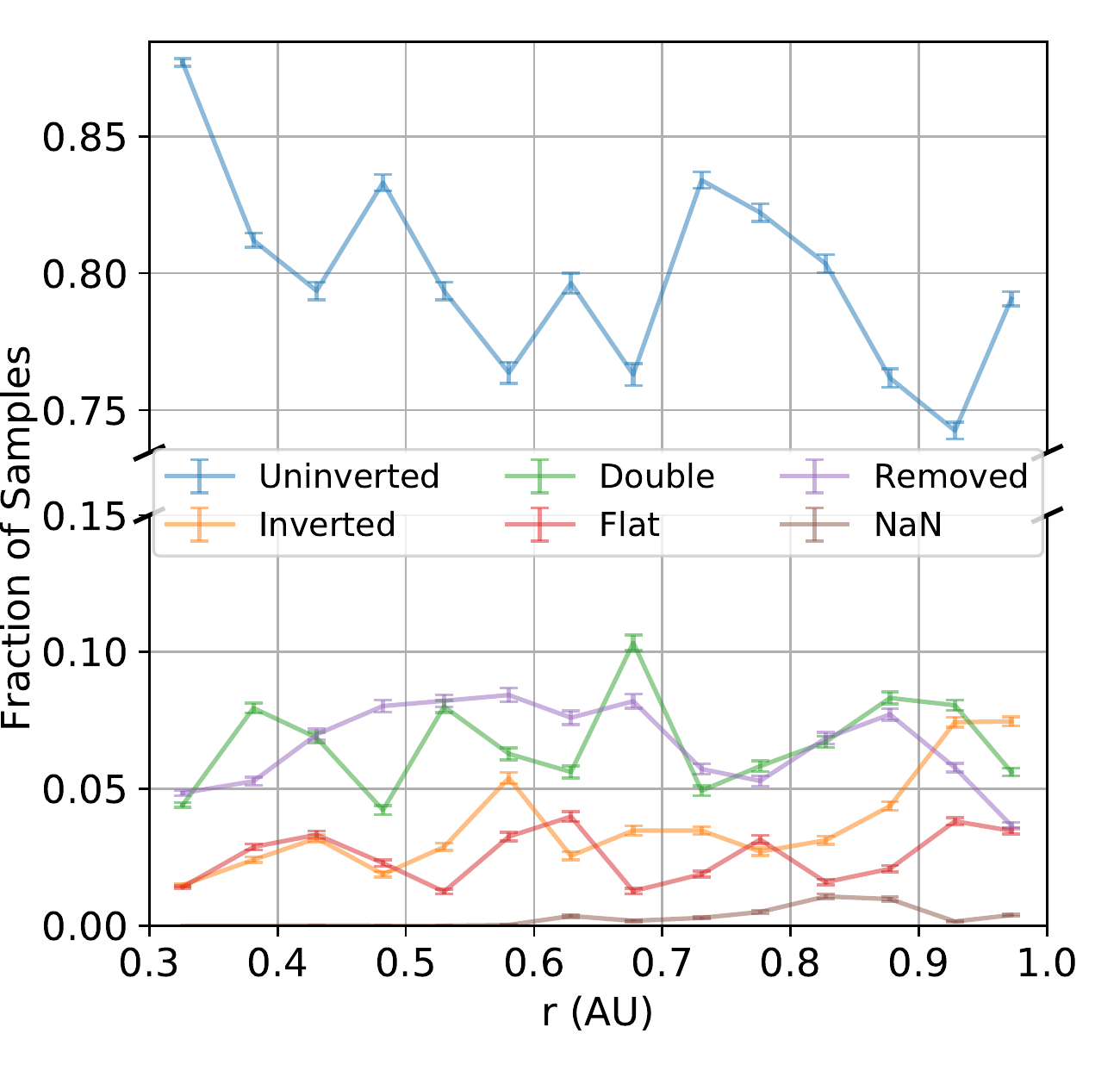}
    \caption{Relative occurrence of 6 classes of HMF/strahl sample against heliocentric distance, $r$, in 14 distance bins. The 4 HMF classes shown in Figure \ref{fig:mainres} are shown alongside the occurrence of samples which are not used in the main portion of the analysis. `Removed' refers to samples with anomalous no-strahl VDFs (Section \ref{sec:datmeth}). `NaN' refers to data in which there are invalid values in either the electron, HMF, or the  ion bulk velocity data. 
    }
    \label{fig:fixed}
\end{figure}
In Section \ref{sec:datmeth} we described data which were discarded from the study because a valid HMF/strahl classification was not possible. In Figure \ref{fig:fixed} we plot the occurrence of the 4 classified valid HMF/strahl types from Section \ref{sec:results}, and additionally 2 types of unclassifiable sample; `removed' which contain the anomalous strahl VDFs, and `NaN' samples for which some crucial data (the electron VDFs, HMF components, or bulk velocity) are missing. The occurrences of the 4 valid  types differ slightly from those in Figure \ref{fig:mainres}, because these correspond to all samples, and not only the valid ones. The `NaN' samples  have very low occurrence, which is concentrated at $r>\SI{0.6}{AU}$. Anomalous strahl occurrence is around 0.05--0.08; comparable to the bidirectional strahl samples, and greater than inverted or flat samples. There does not appear to be a strong radial trend in anomalous strahl occurrence, although there are minima at perihelion and aphelion. 

The occurrence of NaN values is low enough that we are confident that they do not have a significant impact on the primary results of the study. The occurrence of anomalous strahl, which we do not know the origins of, is sufficiently large that we have to consider more carefully. 
Regardless of whether the anomalous strahl is an instrumental/data artefact, or a true solar wind electron phenomenon, there is minimal impact on the results of this study if it is equally likely to occur for all 4 valid HMF types.
The lack of strong radial trend in anomalous strahl occurrence suggests that this is the case, as it does not vary proportionally with any one HMF type in particular.

\begin{figure*}
    \centering
    \includegraphics[width=.7\textwidth]{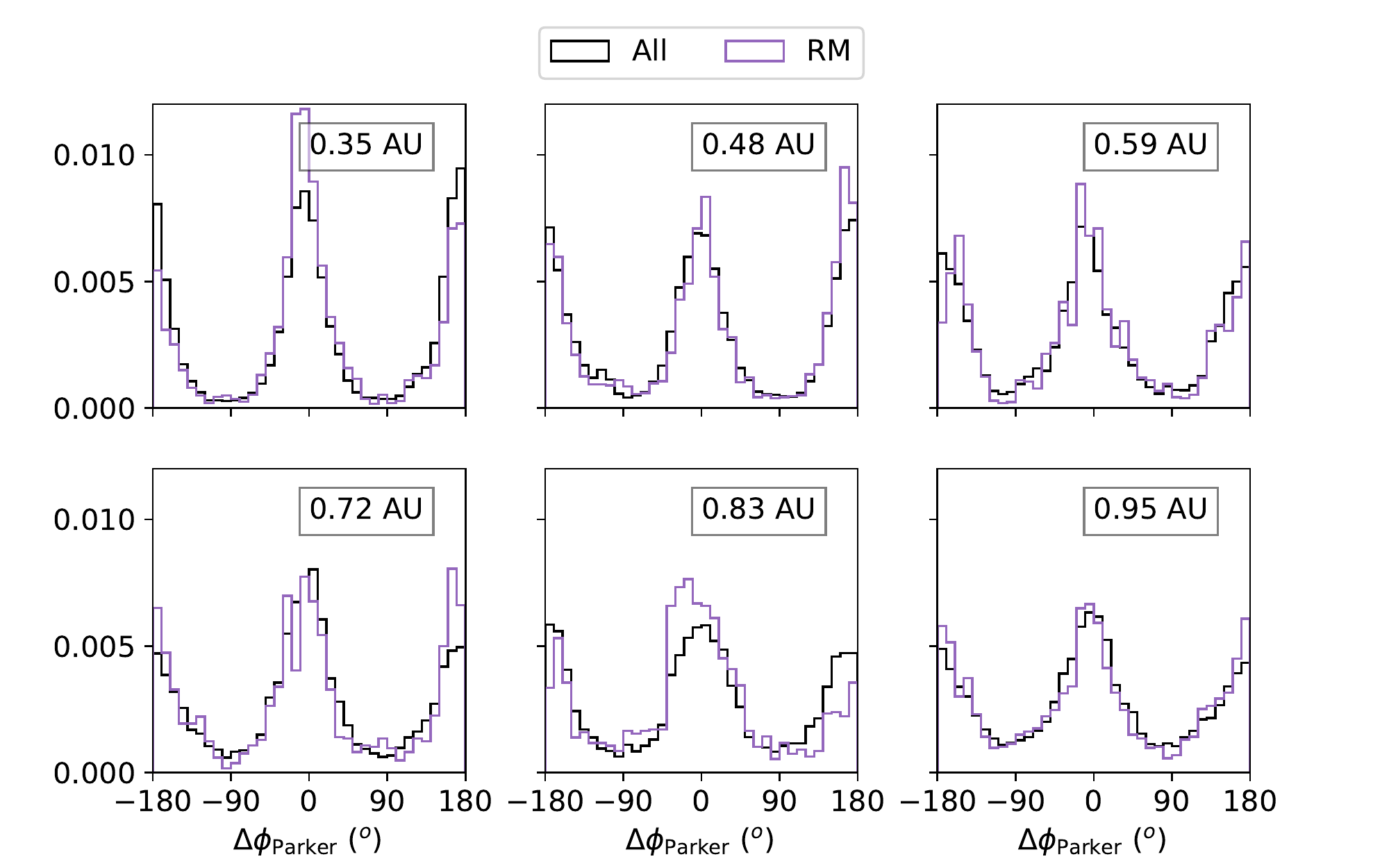}
    \caption{Normalised histograms of HMF angle relative to the nominal Parker spiral direction, $\Delta\phi_{\mathrm{P}}$, in 6 radial distance bins, for all samples in black, and for only those which correspond to anomalous strahl in purple.}
    \label{fig:parkerhist_rm}
\end{figure*}

Figure \ref{fig:parkerhist_rm} shows normalised histograms of HMF angles associated with the anomalous strahl samples in comparison to the HMF angles across data from all HMF types combined. 
The HMF angles of the anomalous samples match reasonably well with those of the combined data; forming peaks centred around \SI{0}{\degree} and \SI{180}{\degree}. The data are noisier, and in some bins more concentrated on one polarity than the other, which is consistent with the low number of samples. Only the distance bin centred on \SI{0.83}{AU} departs notably from the others, as one peak appears skewed away from \SI{0}{\degree}. This may again be caused by sampling effects.
The HMF angle agreement between anomalous strahl and the combined data, suggests that the anomalous strahl samples are not associated with any one particular HMF type.
Thus, their presence in the data, and exclusion from this study, is unlikely to significantly affect the results for the 4 HMF types under consideration.

\section{Heliocentric Distance Bins and Occurrence Errors}\label{sub:bins}
\begin{figure*}
    \centering
    \includegraphics[trim={9.5cm 0 0 0},clip,width = 0.6\textwidth]{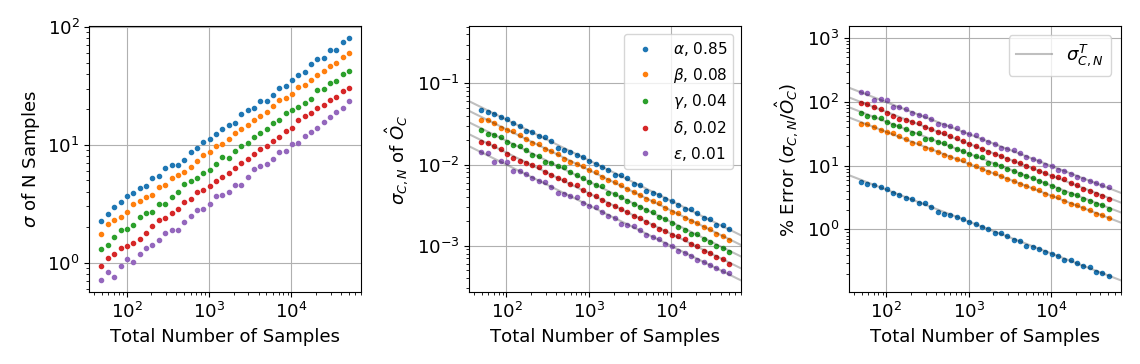}
    \caption{Plots illustrating the errors associated with estimating the proportions of each class within an underlying distribution made up of 5 discrete classes; $\alpha, \beta, \gamma, \delta,$ and $\epsilon$, when sampling that distribution with a variable number of samples $N$. Each class has a true occurrence rate in the underlying distribution, $O_C$, indicated in the legend.  
    \textbf{Left:} Theoretical standard deviations, $\sigma^T_{C,N}$, as calculated from Equation \ref{eq:std} are plotted as solid lines  against $N$. Numerically-derived standard deviations, $\sigma_{C,N}$, in the estimators of the occurrence of each class are overlaid as coloured points.  \textbf{Right:} Corresponding percentage error in estimating each occurrence calculated by dividing $\sigma_{C,N}$ by the true underlying occurrence, $O_C$.}
    \label{fig:errmodel}
\end{figure*}

We wish to determine the minimum required number of data points per heliocentric distance bin, in order to keep errors in occurrence estimates below an acceptable (or at least, known) threshold.  To do so, we estimate  errors in the measured occurrence rate of  predetermined HMF/strahl classes (labelled $\alpha, \beta, \gamma, \delta,$ and $\epsilon$) as a function of the total number of samples, $N$.  The occurrences of each class, $O_C$, are chosen to mimic those anticipated for the valid HMF morphology classes (i.e., one class constitutes $\sim\SI{85}{\percent}$ of the samples, corresponding to uninverted HMF). 
Assuming a binomial distribution, the theoretical standard deviation $\sigma^T_{C,N}$, in the estimator of the occurrence of each class, $\hat{O}_C$, given $N$ total samples is
\begin{equation}\label{eq:std}
    \sigma^T_{C,N} = \sqrt{\frac{P\{C\}(1-P\{C\})}{N}}
\end{equation}
where $P$ is the probability of sampling the class $C$, which is equivalent to the occurrence of that class in the underlying distribution. The binomial distribution is appropriate here despite having $>2$ classes, as for each class $C$ we can consider the underlying distribution to be a binomial, where the values are either $=C$ or $\neq C$.

The above analytical error can also be tested on synthetic data to confirm that our description of error as a function of bin size (and therefore our choice of bin size) is appropriate. Figure \ref{fig:errmodel} plots the standard deviation results derived from Equation \ref{eq:std} against $N$. It also shows the  standard deviation from a numerical Monte Carlo sampling simulation, $ \sigma_{C,N}$, for the same underlying distributions for verification of the method.
There is strong agreement between the numerical and analytical estimates of this error, confirming that the binomial error estimate is appropriate. The largest standard deviation applies for $\alpha$, which makes up \SI{85}{\percent} of the underlying distribution, while the smallest applies for $\epsilon$, which makes up \SI{1}{\percent}. However, we require an acceptable maximum relative error to apply to each measured occurrence, and so in the right panel we plot $\sigma^T_{C,N}/O_C$, which gives the percentage error. The largest percentage error is found for the \SI{1}{\percent} occurrence class, $\epsilon$. An acceptable percentage error of \SI{10}{\percent} in occurrence rate (for classes with occurrence rate $\geq0.01$) corresponds to $N=\SI{e4}{}$ samples in each bin.

We note here that using this error estimate for our study implicitly assumes that the sampling of the underlying distribution of HMF morphologies by  \textit{Helios} 1 is truly random, and that our classification procedure in Section \ref{sec:dis} is also unbiased. We  consider if this is the case in Section \ref{sec:dis}. 


\bsp	
\label{lastpage}
\end{document}